\documentclass[12pt]{article}
\pdfoutput=1
\usepackage[a4paper]{geometry}
\usepackage{jheppub, amsmath,amssymb,amsfonts,amsxtra, mathrsfs, makeidx,graphics,graphicx,amsthm,epsfig, bm,longtable,float, color,tikz,mathtools,xfrac,footnote,rotating, lscape, makecell, environ,mathtools, empheq}

\usepackage{subfig}

\usepackage{amsthm}
\usepackage{hyperref}

\usepackage{amsmath}
\usepackage{amsfonts}
\usepackage{commath}
\usepackage[english]{babel}
\usepackage{setspace}

\usepackage{epsfig,amssymb} 
\usepackage{amsmath}

\usepackage{amscd,color}
\usepackage{amsmath,graphicx}
\usepackage{verbatim}

\usepackage{latexsym}
\usepackage{amsmath,amsfonts,amssymb,amsthm}
\usepackage{amsmath,amsthm}

\title{
\begin{center}
Analytic bounces in {\bf d} dimensions
\end{center}}

\author[a]{Antonio Amariti}
\affiliation[a]{INFN, Sezione di Milano, Via Celoria 16, I-20133 Milano, Italy}

\emailAdd{antonio.amariti@mi.infn.it}

\abstract{We study the Euclidean bounce action 
interpolating  between a false and a true vacuum
for a scalar field theory with various types of potential.
We focus on the cases of a triangular, a square and a quadratic barrier,
where the bounce action has already been computed analytically in four dimensions.
We generalize the result to $d$ dimensions, providing an analytic formula in each case.
Furthermore we show that our results reduce to the ones computed 
from the thin wall approximation, when the true and the false vacuum are close in energy.
When the true vacuum cannot be reached in a finite amount of Euclidean time we 
study the damped oscillations of the solution by analytical continuation to Lorentzian spacetime.
}

\begin{document}

\maketitle

\section{Introduction}

The scalar potential $V(\phi)$ of a generic quantum field theory can allow for the existence of
multiple vacua. In examples with two 
non degenerate vacua we can distinguish between a true vacuum $V(\phi_-) = V_-$  and a false vacuum  $V(\phi_+) = V_+$ .
The false vacuum state is unstable and it is expected to decay via tunneling \cite{Kobzarev:1974cp}.
The original approach to the study of the decay of a false vacuum  was carried on by Coleman in a 
series of papers \cite{Coleman:1977th,Callan:1977pt,Coleman:1977py,Coleman:1980aw}
by studying the nucleation of a true vacuum bubble inside a false vacuum.
The problem was reformulated in terms of the evaluation of the
action of the bounce solution of the equations of motion  
at Euclidean time $r$.
In general the important condition beyond the existence of such a bounce solution 
corresponds to have a correct choice for the initial value for the field in the inverted potential, dictated by a calibration of
undershooting and overshooting in order to fulfill the requirement that $\phi(r) \rightarrow  \phi_+$ if $r \rightarrow \infty$.

This general picture can be made quantitative by studying explicit potentials.
In \cite{Coleman:1977py} it was studied the thin wall approximation, i.e. a case with two almost degenerate vacua 
in which the computation of the bounce action was not necessary for the evaluation of the tunneling rate.
However if this approximation is not valid one needs to compute the bounce action from the functional form of the potential.
Even if numerically one can use shooting methods \cite{Adams:1993zs,Sarid:1998sn} or more sophisticated techniques
\cite{Espinosa:2018hue,Jinno:2018dek,Guada:2018jek,Guada:2019roh,Guada:2020xnz}
 to approach the problem, it is interesting to study cases where an analytic solution 
can be obtained.
For example, in 4d the cases of a triangular and a square potential barrier have been studied in detail in \cite{Duncan:1992ai}.
Even if these cases may look unphysical\footnote{See \cite{Dutta:2012qt} for a discussion on the validity of the triangular approximation.}, 
due to their singular behavior, they have many physical 
features. For example the triangular approximation has been used to estimate the lifetime of 
metastable supersymmetry breaking vacua \cite{Intriligator:2006dd}.
Non singular 4d quadratic potentials have been studied in \cite{Pastras:2011zr,Dutta:2011ej}
while  cubic and quartic cases have been further discussed in \cite{Dutta:2011fe,Dutta:2011rc}.

There are in general two possibile behaviors of the solution that interpolates between the two vacua: 
either the field reaches the true vacuum at finite Euclidean time or it does not.
In the first case the analytic continuation of the solution is straightforward while in the second case more care is needed.
In the triangular and in the quadratic case such a continuation has been studied in 4d in \cite{Pastras:2011zr}, where it has
been shown that, if  the parameters of the potential do not allow to reach the true vacuum at finite Euclidean time, then 
there are damped oscillation around the true vacuum inside the bubble, once the analytic continuation is considered.

\begin{figure}
\begin{center}
\begin{tabular}{cc}
\includegraphics[width=7cm]{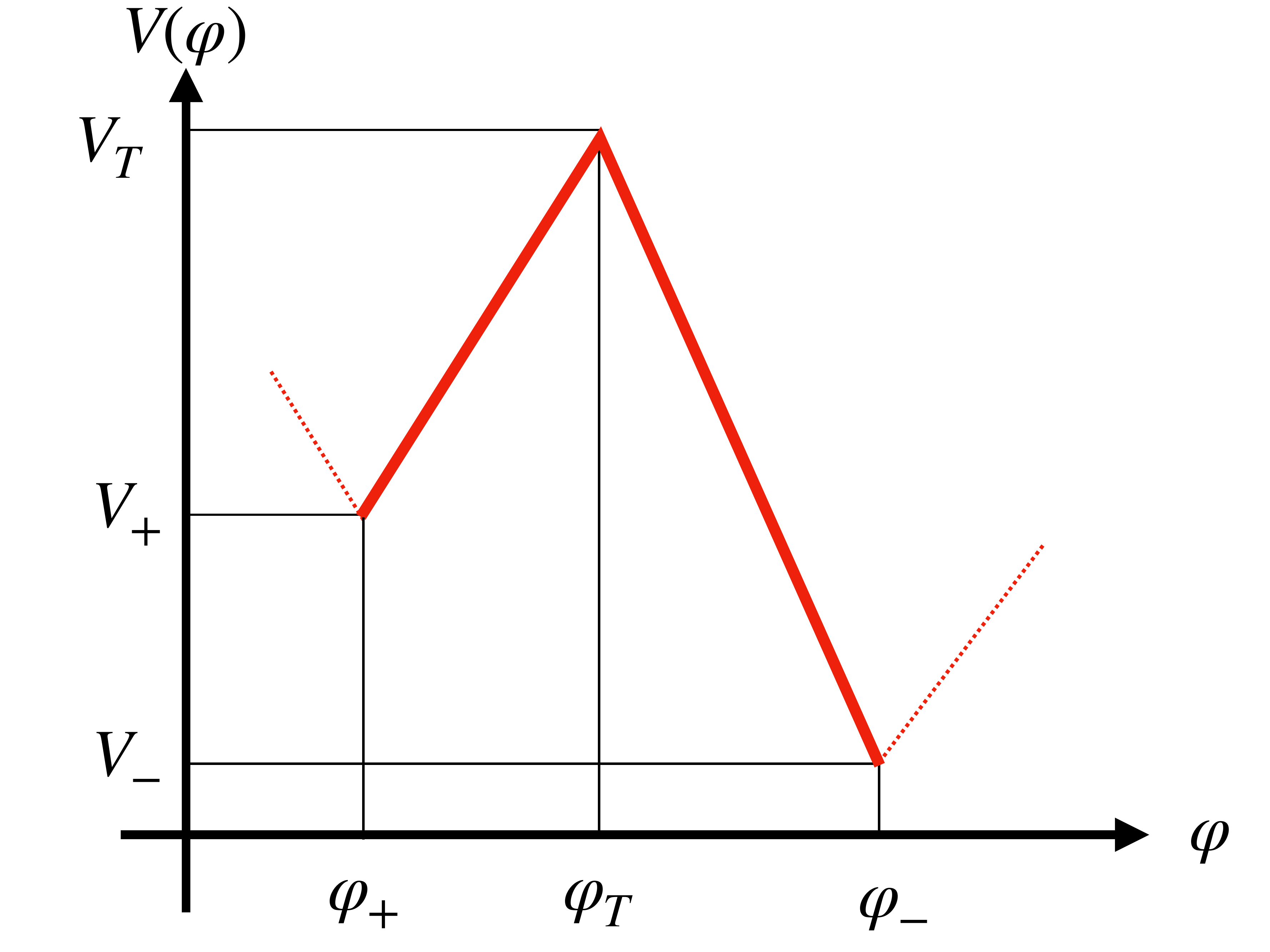}
&
\includegraphics[width=7cm]{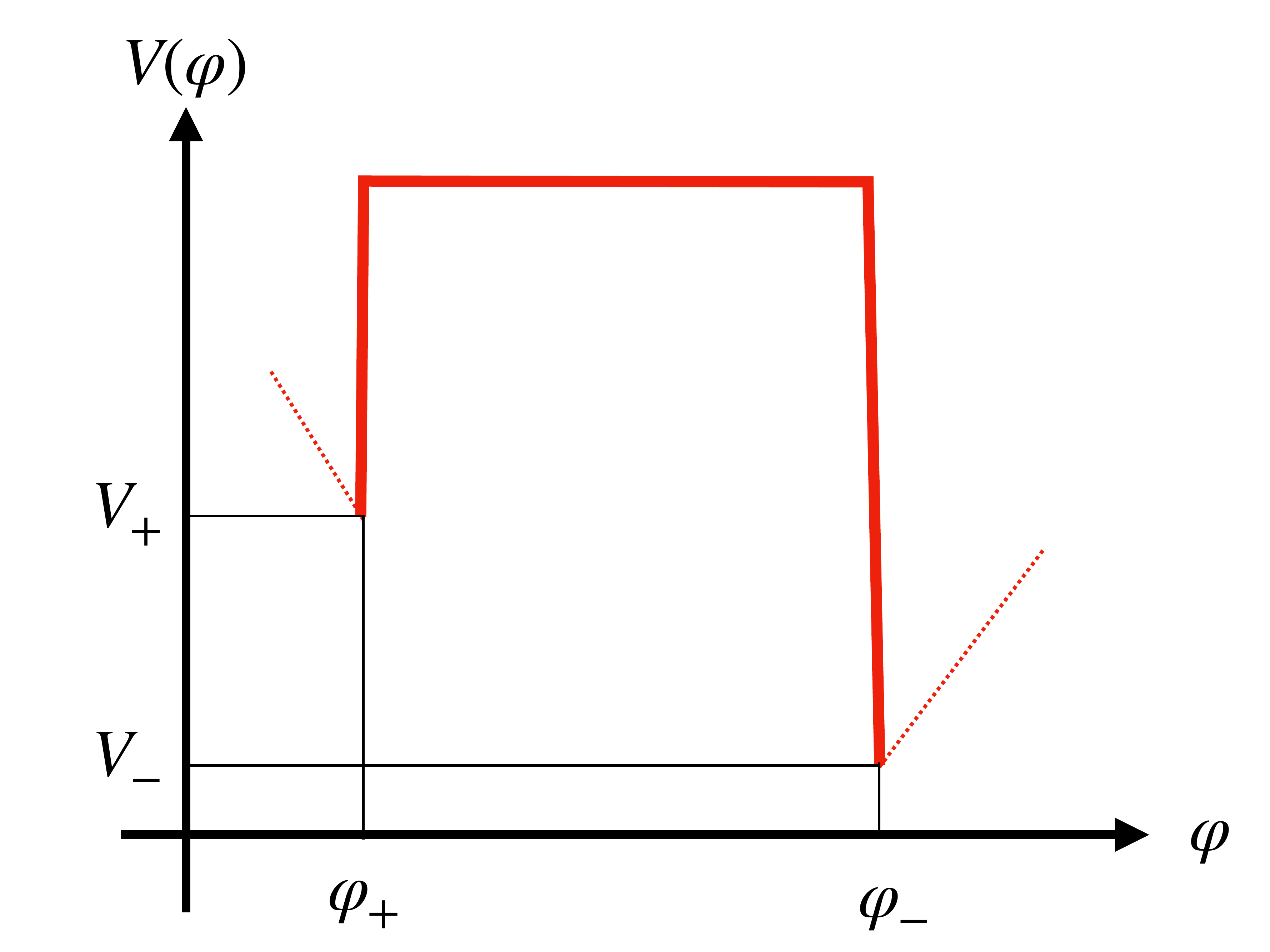}
\\
\includegraphics[width=7cm]{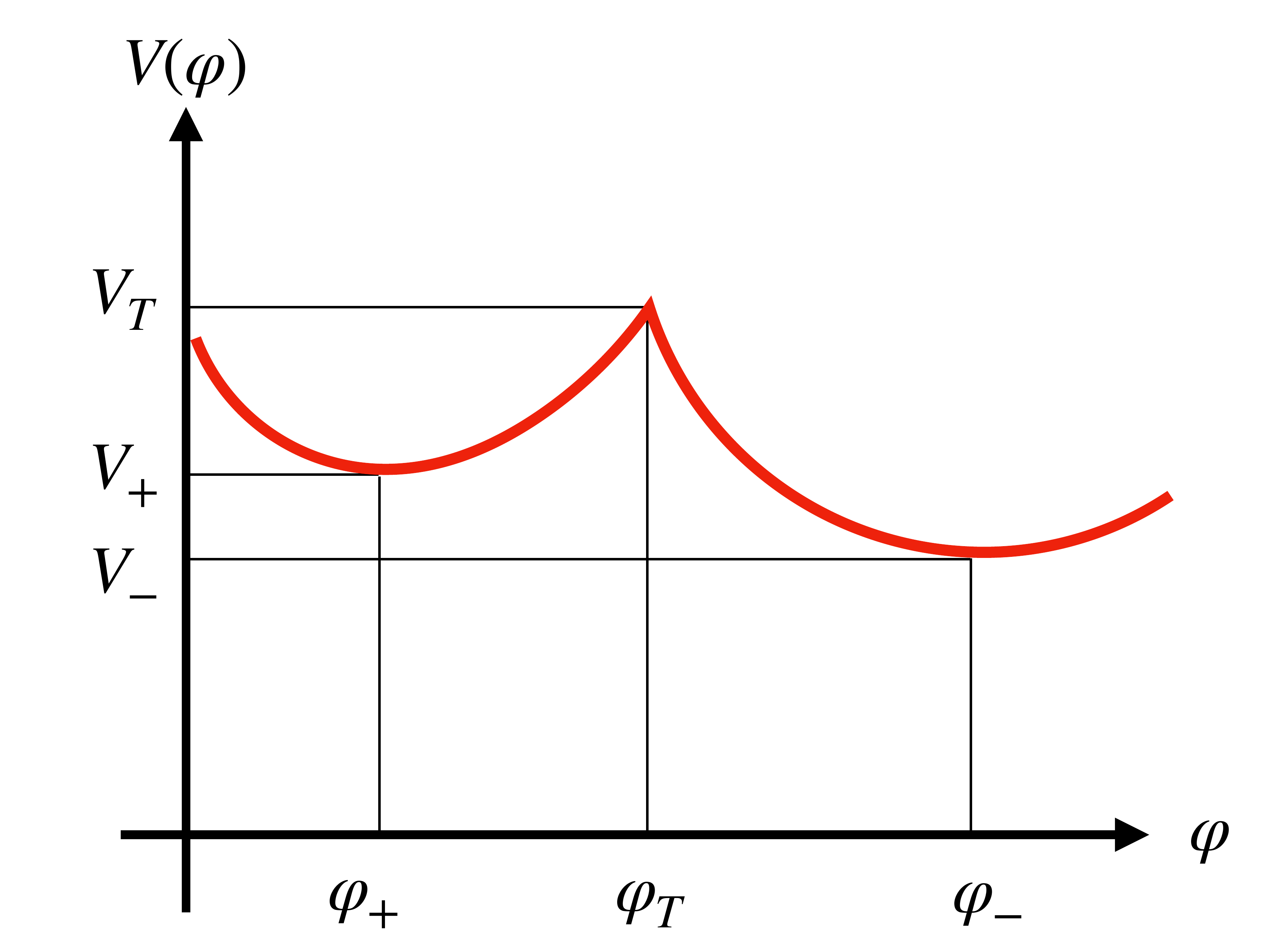}
&
\includegraphics[width=7cm]{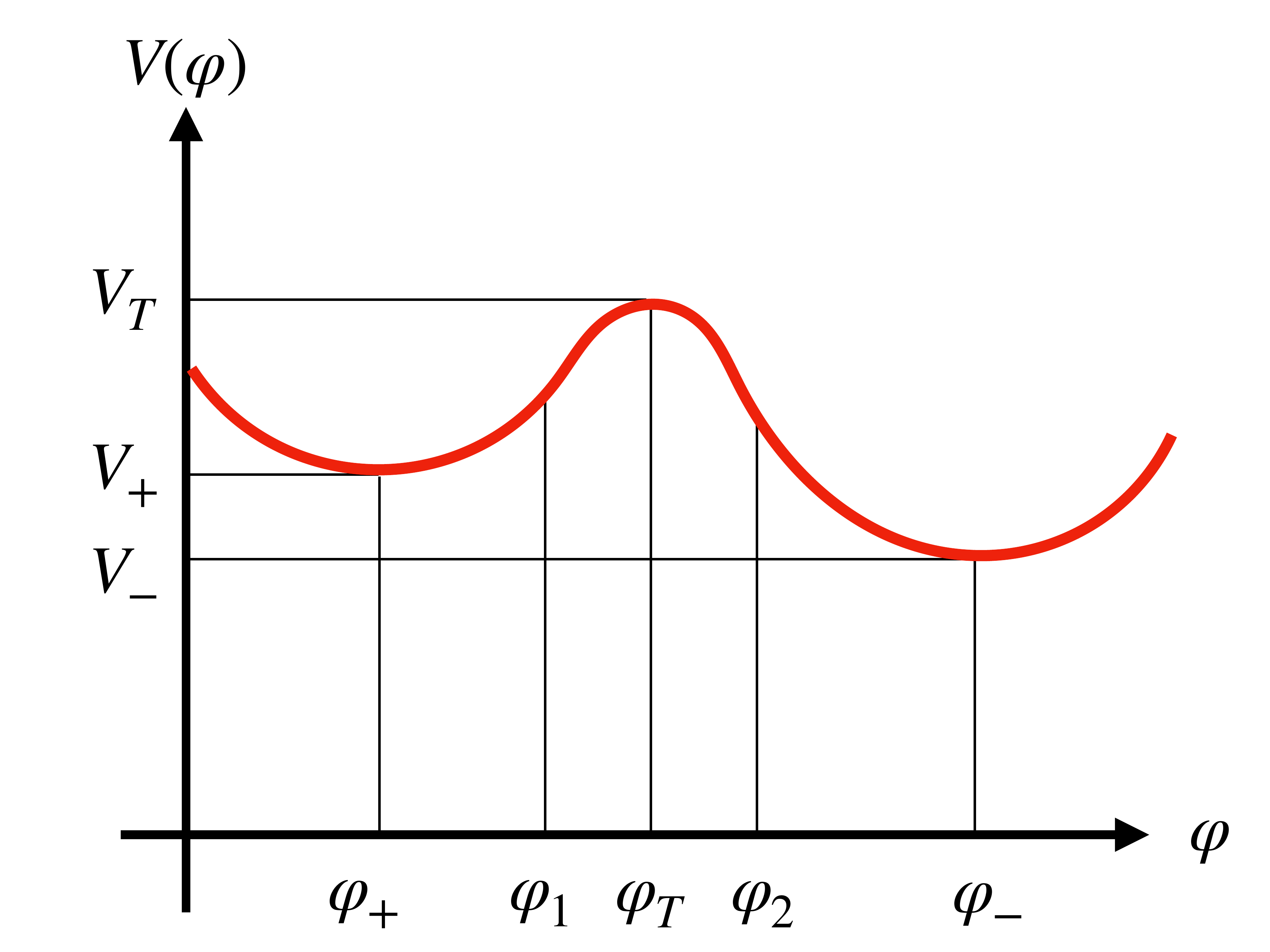}
\end{tabular}
\caption{The scalar potentials studied in this paper. The first potential corresponds to the triangular barrier while the second one 
is the square barrier. They have been analyzed in \cite{Duncan:1992ai}  in the 4d case. The third potential and the fourth one have been studied in 
4d in \cite{Pastras:2011zr}. The former  has been referred in \cite{Pastras:2011zr} as the volcanic potential. It consists of a piecewise potential, quadratic  at $V_{\pm}$ 
with a cusp at $V_T$. In the last case the cusp at $V_T$ is removed and it is replaced by  a quadratic cap.} 
\label{fig0}
\end{center}
\end{figure}

Motivated by this series of exact results in this paper we study the bounce action for a linear,  a square
and a quadratic potential in $d$ dimensions.
In Figure \ref{fig0} we provide a schematic picture of the various potentials discussed in this paper.
In all these cases the equations of motion have an exact solution and one can use these solutions to 
compute the bounce action analytically.
In our analysis we observe that many of the physical issues of the 4d case can be extended to $d$ dimensions 
offering a  complete and unifying picture. For example we can reduce our results in the various cases to the 
thin wall approximation and, when the vacua are not reached at finite Euclidean time,  
we find the damped oscillations in the analytic continuation of our solutions.

The paper is organized as follows. In section \ref{secga} we collect the various results necessary for the analysis of the bounce action 
in $d$ dimensions. In section \ref{sectri} we study the triangular potential distinguishing  two different behaviors, depending on the possibility to reach or not the true 
vacuum at finite Euclidean time. We discuss also the relation with the thin wall and the existence of damped oscillations 
in the analytic continuation of the solution inside the bubble. 
In section \ref{secsqu} we study the bounce solution for a triangular barrier in $d$ dimensions and we discuss the thin wall limit of our formula, matching it with the expected result.
In section \ref{secqu} we study the quadratic potential. We distinguish two cases: in the first case
the potential consists of two branches with  a  quadratic behavior connected by a cusp at the local maximum while in the second case this cusp is removed in favor of a smooth quadratic cap.
In section \ref{secon} we conclude.

\section{General aspects}
\label{secga}
When considering a scalar field theory with a single field and a false and a true vacuum 
the analysis of the tunneling solution interpolating between them can be obtained by considering spherically symmetric solutions \cite{Coleman:1977th}
and reducing the problem to one dimension 
considering the Euclidean time (or radius) $r = \sqrt{t^2+|\vec x|^2}$.
In this case the $d$ dimensional action becomes
\begin{equation}
\label{action}
S_E [\phi] = \frac{2 \pi ^{d/2}}{\Gamma \left(\frac{d}{2}\right)}
\int_0^\infty {\bf d} r \, r^{d-1} \left(\frac{1}{2} {\phi'}^2
+ V(\phi)
\right)
\end{equation}
and the relative equation of motion is 
\begin{equation}
\label{eom}
\frac{\partial^2 \phi(r)}{\partial r^2} + \frac{d-1}{r} \frac{\partial \phi(r)}{\partial r} =V'(\phi(r))'
\end{equation}
The boundary conditions that have to be imposed are  
\begin{equation}
\label{bcx}
\lim_{r \rightarrow \infty} \phi(r) = \phi_+ \quad \&\quad
\frac{\partial \phi(r)}{\partial r} \Big|_{r=0} = 0
\end{equation}
where $\phi_+$ corresponds to the field at the false vacuum.
Once the solution is known one can compute the decay rate of the false vacuum into 
the true one 
\begin{equation}
\Gamma = A e^{\frac{B}{\hslash}}(1+o(\epsilon))
\end{equation}
where the $d$-dimensional bounce $B$ is obtained by evaluating the action along the solution and
then subtracting the action at the false vacuum
\begin{equation}
\label{SB}
B= S_E(\phi(r))-S_E(\phi_+) 
\end{equation}
  For a generic potential in $d$-dimensions this procedure cannot be performed analytically and 
 numerical techniques are needed.
 Nevertheless, for specific choices of the potential, the equation (\ref{eom}) can be solved
 and the analytic form of the solution $\phi(r)$ can be used to estimate the bounce (\ref{SB}).

 In some cases, when $\Delta V_- - \Delta V_+$ is very small with respect of 
 the scales of the theory, one can use the thin wall approximation that does not require the 
 explicit evaluation of the integral. 
 In a generic $d$-dimensional setup the bounce action arising from the thin wall approximation
 is 
 \begin{equation}
 \label{bbbb}
 B_{t.w.}=
 \frac{2 \, \pi ^{d/2} }{d \, \Gamma \left(\frac{d}{2}\right)}  \left(\frac{(d-1) }{\epsilon }\right)^{d-1} S_1^d
\end{equation}
where
\begin{equation}
\label{s1s1}
S_1 = \int_{\phi_+}^{\phi_-} d\phi \sqrt{V(\phi) -V(\phi_+)}
\end{equation}

 %
 %
 %
 %
 %
 %%%%%%%%%%%%%%%
 %%%%%%%%%%%%%%%
\section{The triangular barrier}
\label{sectri}
 %%%%%%%%%%%%%%%
 %%%%%%%%%%%%%%%
 %
 %
 %
 %
 %
We start our analysis by studying the case of a triangular barrier in $d$-dimensions.
The scalar potential connecting the false vacuum and the true vacuum is
\begin{equation}
\label{pote}
V(\phi) = \left \{ 
\begin{array}{cc}
\phantom{-} \lambda_{+} (\phi-\phi_+) + V_+ & \quad \phi<\phi_T 
\vspace{.1cm}\\
- \lambda_{-} (\phi-\phi_-) + V_- &\quad \phi>\phi_T \\
\end{array}
\right.
\end{equation}
In these expressions $\phi_T$ represents the field at the maximum, with $V(\phi_T) \equiv V_T$.
Keeping the same notations of \cite{Duncan:1992ai} we also define $c \equiv \frac{\lambda_{-}}{\lambda_{+}}$ and
$\lambda_{\pm }  =  \frac{\Delta V_\pm }{\Delta \phi_\pm} ,$
with $\Delta V_\pm = (V_T - V_\pm)$ and $\Delta \phi_\pm = \pm(\phi_T - \phi_\pm)$.
 
Next we have to study the solution $\phi(r)$ interpolating between the two vacua in the Euclidean time $r$
by solving the equation of motion (\ref{eom}) in the two branches $\phi<\phi_T$ and $\phi>\phi_T$.
This is done by providing opportune boundary conditions and by requiring that the solution is smooth 
at $\phi = \phi_T$.
This provides a set of equations that allow us to express the values of $r=R_+$,  $r=R_T$ and $r=R_-$ 
in terms of the parameters of the potential.
There are two types of boundary conditions that we have to impose, depending on the fact that the 
field can either reach or not the true vacuum at finite Euclidean time. We distinguish these two cases in the following.
Once the solution interpolating the two vacua is found we can plug it in the bounce action (\ref{SB})
and to estimate the lifetime of the false vacuum, i.e. the decay rate of this state in the true vacuum.

\subsection{First case, $\phi_0 < \phi_-$}
\label{firstTri}

In this case the solution does not reach the true vacuum at finite Euclidean time.
The first boundary condition, that has to be satisfied by both the 
cases is that the field reaches the false vacuum at finite radius $R_+$ and stays there:
\begin{equation}
\label{bcf}
\phi(R_+) = \phi_+,\quad \frac{\partial \phi(r)}{\partial r} \Big |_{r=R_+} = 0
\end{equation}
The second boundary condition is related to the initial value of the field $\phi$
at $r=0$.
In the first case we study the situation where  $\phi(0) = \phi_0<\phi_-$,
with 
$\frac{\partial \phi(r)}{\partial r} \Big |_{r=0} = 0$.
The second case, when the field stays at $\phi_-$ for a certain amount of Euclidean time $R_0$ 
will be studied in subsection \ref{secondTri}.

\begin{figure}
\begin{center}
\includegraphics[width=10cm]{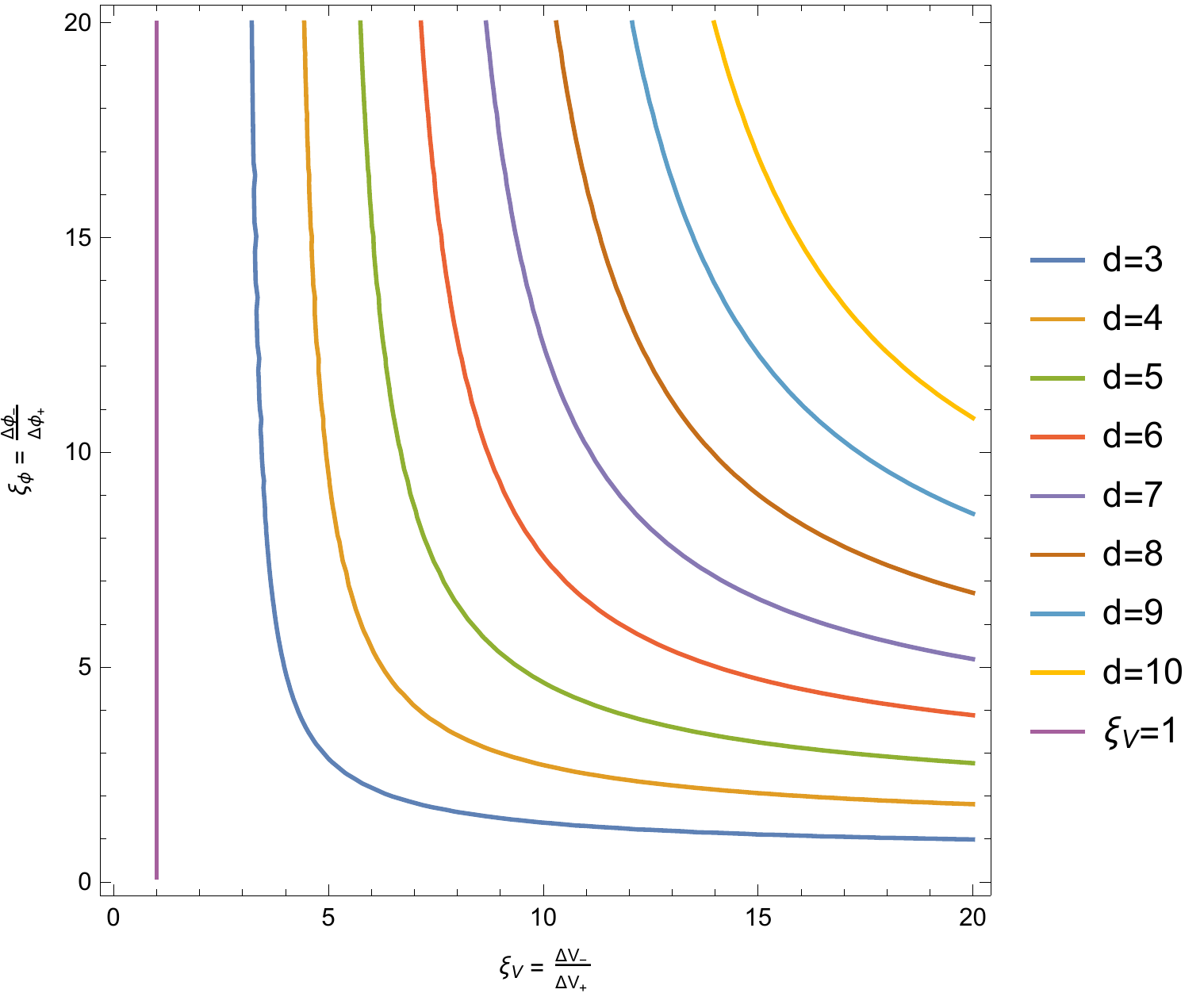}
\caption{Graphical solution of the equation
$ \xi_\phi \big(2 \xi_V  + d \, \xi_\phi  \big(1-\big(\frac{\xi_\phi +\xi_V  }{\xi_\phi }\big)^{2/d}\big)\big)=(d-2) \xi_V $ for various values of the spacetime dimension $d$. 
For each value of $d$  the inequality (\ref{ineq}) holds is the region above the curve. Below the curve we look for solutions of the first case, \emph{i.e.} when the true vacuum can be reached in a finite amount of Euclidean time, while 
above each curve we are in the second case,  \emph{i.e.} when the vacuum cannot be reached in a finite amount of Euclidean time. In the figure we have also plotted the line corresponding to the ratio $\frac{\Delta V_+} {\Delta V_-}=1$.
This is the limit when the thin wall approximation can be used. We observe that such approximation is possible only in the second case.}
\label{fig1}
\end{center}
\end{figure}

The solution of the equation of motion (\ref{eom}) along the two branches of the potential (\ref{pote}) 
can be separated into a solution for $0<r<R_T$, where $R_T$ is an unknown to determine and into
a solution for $R_T<r<R_+$. We find
\begin{equation}
\phi(r) = \left\{
\begin{array}{lc}
\phi_R(r) = \phi_0-\frac{\lambda_{-}}{2d} r^2 &\quad r<R_T \\
\phi_L(r)  = \phi_+ + \frac{\lambda_{+} \left( (d-2)r^d - d \, r^{d-2}R_+^2 +2 R_+^d\right)}{2d(d-2)r^{d-2}} 
&\quad R_T<r<R_+
\end{array}
\right.
\end{equation}
By imposing $\phi_R(R_T) = \phi_L(R_T) = \phi_T$ and the $\phi_R'(r) \Big |_{r=R_T} = \phi_L'(r) \Big |_{r=R_T} $ we find the following relations
\begin{eqnarray}
\phi_0 &=& \phi_{T} + \frac{\lambda_{-}}{2d} R_{T}^2
\nonumber \\
\Delta \phi_+ &=&\lambda_{+}  \frac{2c-d((c+1)^\frac{2}{d}-1)}{2 d (d-2)} R_{T}^2
\\
R_+ &=& (c+1)^{\frac{1}{d}} R_{T}
\nonumber
\end{eqnarray}
These three equations allow us to determine $R_+$, $R_T$ and $\phi_0$ in terms of the parameters 
of the potential.
Eventually we  compute the bounce action along this solution and we obtain
\begin{eqnarray}
S_B =\frac{4 (c+1)}{d (d+2) \Gamma \left(\frac{d}{2}\right)}
 \left(\frac{2 \pi  (d-2) d}{2c-d((c+1)^{\frac{2}{d}}-1)}\right)^{\frac{d}{2}}
\frac{ \Delta \phi _+^d}{\Delta V_+^{\frac{d-2}{2}}}
\end{eqnarray}
As discussed above this analysis is valid if the solution has not yet reached the true vacuum $\phi_-$, 
i.e. for $\phi_0\leq \phi_-$. This is equivalent to the condition 
\begin{eqnarray}
\label{notyet}
\frac{\Delta \phi_-}{\Delta \phi_+}
> \frac{c (d-2)}{2c-d \left((c+1)^\frac{2}{d}-1\right)}
\end{eqnarray}
Defining  $\xi_V \equiv \frac{\Delta V_-}{\Delta V_+} >0$ and
 $\xi_\phi \equiv \frac{\Delta \phi_-}{\Delta \phi_+} >0$ the inequality  (\ref{notyet}) can be written as
\begin{equation}
\label{ineq}
 \xi_\phi >\frac{(d-2) \xi_V }{2 \xi_V + d \, \xi_\phi  \left(1-\left(\frac{\xi_\phi +\xi_V }{\xi_\phi }\right)^{2/d}\right)}
 \end{equation}
This inequality cannot be studied analytically for generic $d$, so we  study it graphically in Figure \ref{fig1}:
the region above each curve represents the region where (\ref{ineq}) is satisfied. 
If we are in the other region, below the curve,  then $\phi_0 > \phi_-$ and we must modify the 
choice of the boundary condition  close to the true vacuum.
This case will be discussed in the next subsection.

\subsection{Second case, $\phi_0 > \phi_-$}
\label{secondTri}

In this case a solution can be obtained if we force the field to stay at the true vacuum $\phi_-$ for a certain amount of Euclidean time $R_-$.
After that the dynamics is driven by the equation of motion (\ref{eom}) with the inverted
potential and the field can reach the false vacuum $\phi_+$ at Euclidean time  $R_+$.
While we can keep the same boundary conditions (\ref{bcf}) at the false vacuum, the boundary conditions at the true vacuum become $\phi(r) = \phi_-$ for $0<r<R_-$ and 
$\frac{\partial  \phi(r)}{\partial r} \Big|_{r=R_-} = 0$.
The solution to (\ref{eom}) with these boundary condition and with the potential (\ref{pote}) is
\begin{equation}
\left\{
\begin{array}{lc}
\phi_R(r) = \phi_-  & \quad r<R_- \\
\phi_R(r) =\phi _-  + \frac{\lambda _- \left(d \, R_-^2-2 r^{2-d} R_-^d-(d-2) r^2\right)}{2 (d-2) d} &\quad R_-<r<R_+  \\
\phi_L(r)  = \phi_+ -\frac{\lambda _+ \left(d \, R_+^2-2 r^{2-d} R_+^d-(d-2) r^2\right)}{2 (d-2) d} &\quad r>R_+
\end{array}
\right.
\end{equation}
Again we imposing $\phi_R(R_T) = \phi_L(R_T) = \phi_T$ and  $\phi_R'(r) \Big |_{r=R_T} = \phi_L'(r) \Big |_{r=R_T} $ and we find the following relations
\begin{eqnarray}
\label{phiTF}
R_{T }^d=\frac{R_+^d + c R_-^d}{c+1 }, \quad
\Delta \phi _\pm=\frac{\lambda _+}{d-2} \left(\frac{R_\pm^d}{d R_T^{d-2}}+\frac{(d-2) R_{T }^2}{2 d}-\frac{R_\pm^2}{2}\right)
\end{eqnarray}
For a generic choice of spacetime dimension $d$ we cannot solve these equations in terms of the unknowns $R_T$ and $R_\pm$.
For this reason we proceed by computing the bounce action in terms of $R_\pm$, leaving  their dependence on  the parameters of the potential implicit.
In this way we find a very general expression for $B$
\footnote{Observe that the formula here fixes a typo in \cite{Amariti:2009kb} where the role of $R_-$ and $R_+$ was exchanged}:
\begin{equation}
\label{gentriii}
B =.  
\frac{\pi ^{d/2} \lambda _+ \left(\Delta \phi _+ R_+^d-c \Delta \phi _- R_-^d\right)}{\Gamma \left(\frac{d}{2}+2\right)}
\end{equation}
In order to check the validity of this formula we can compare it with the result expected from the thin wall approximation:
\begin{eqnarray}
\label{CTW}
B_{t.w.} = \frac{2^{\frac{1}{2} (3 d+2)} (d-1)^{d-1} \pi ^{d/2} \Delta \phi ^d V_+^{d/2}}{3^d \, d \, \epsilon ^{d-1} \, \Gamma \left(\frac{d}{2}\right)}
\end{eqnarray}
This comparison was done explicitly in the $d=4$ case in \cite{Duncan:1992ai}. Here we can study the $d=3$ and the $d=6$ case 
by solving the equations (\ref{phiTF})  and expanding them in the limit small $\epsilon$ limit, where 
$\Delta V_+ - \Delta V_+ = \epsilon + o (\epsilon^2)$.
These expansions are studied in the appendix \ref{app36}.

From these result we can make an educated guess on the general form of $R_T$ and $R_\pm$ in the thin wall regime
\begin{equation}
\label{thesew}
\begin{array}{l}
R_T=\frac{4(d-1) \Delta \phi  \sqrt{\Delta V_+ }}{3 \sqrt{2} \epsilon} + \beta  + o(\epsilon)  \\
R_\pm = \frac{4(d-1) \Delta \phi \sqrt{\Delta V_+}  }{3 \sqrt{2}\epsilon }\pm\frac{\sqrt{2} \Delta \phi_\pm} {\sqrt{\Delta V_+}}+\beta+o(\epsilon )
\end{array}
\end{equation}
We can then check that these expressions solve (\ref{phiTF}) at the lowest orders in $\epsilon$.
Observe that even if the order $o(\epsilon^0)$ in $R_{T}$ is left unknown we can compute the 
$o(\epsilon^0)$ pieces in $R_\pm$ in terms of $\beta$ and we can then check that this allows us to 
match the expansion of the bounce action computed here with the one expected from the thin wall approximation.
Indeed, substituting (\ref{thesew})   in (\ref{gentriii}), using $\Delta V_- =  \Delta V_+ + \epsilon$ and expanding for small $\epsilon$ we recover the expected formula
(\ref{CTW}) for the bounce action obtained from the thin wall approximation.

\subsubsection{Damped oscillation}
\begin{figure}
\begin{center}
\includegraphics[width=8cm]{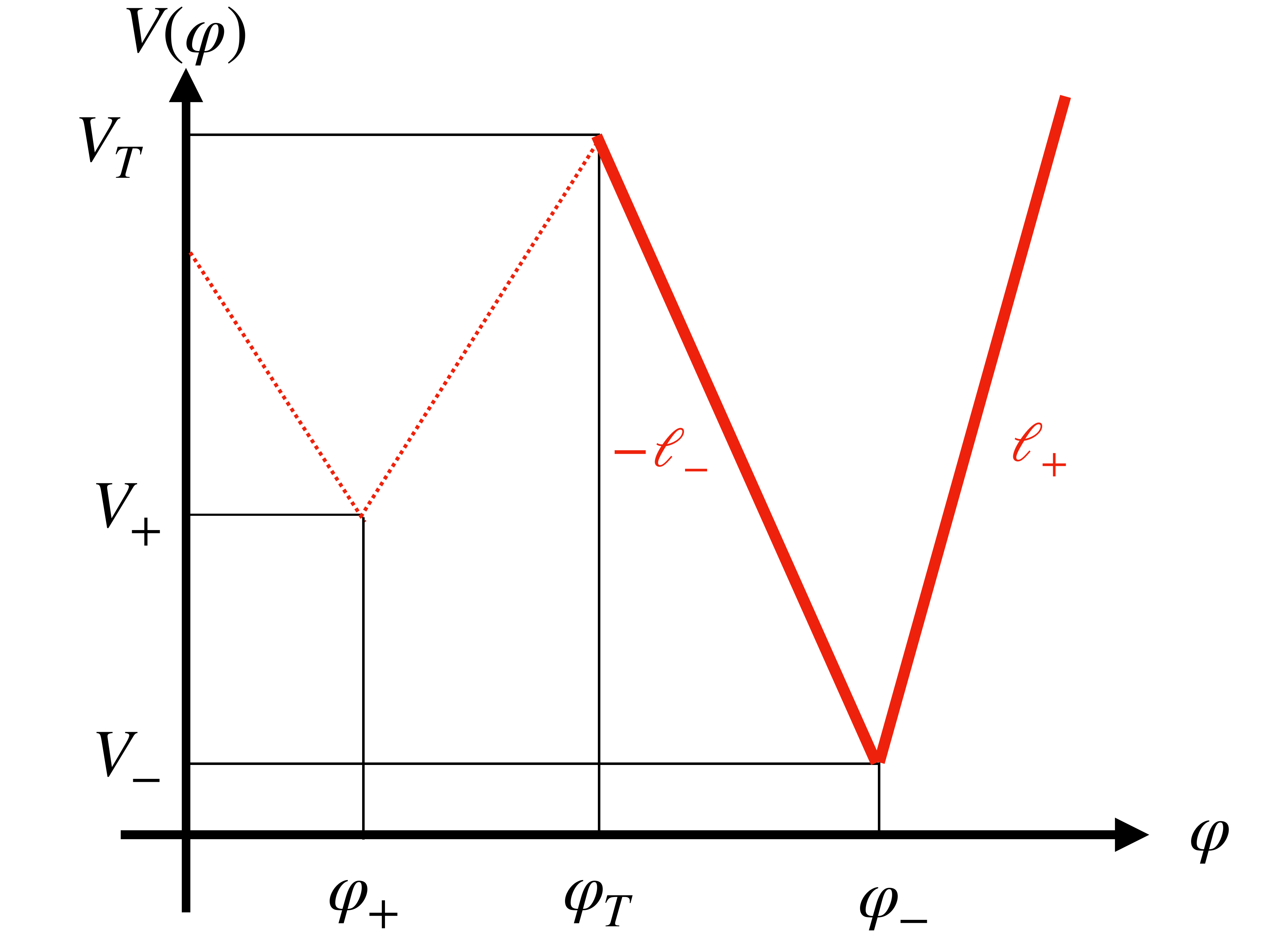}
\caption{The scalar potential around the true vacuum.}
\label{figell}
\end{center}
\end{figure}
Here we study the behavior of the solution if it does not reach the true vacuum.
In this case it has been shown in 4d  \cite{Pastras:2011zr} that the
analytically continued solution asymptotically approaches the vacuum
with an oscillatory behavior.
We expect a similar behavior in  $d$-dimensions. We then analytically continue the Euclidean time $r \rightarrow  i \tau$
and, in order to simplify the notations, we refer to the derivatives of the potential around the vacuum as $\pm \ell_{\pm}$
as in Figure \ref{figell}.
The relevant  part of the scalar potential around the true vacuum is
\begin{equation}
V(\phi)=
\left\{
\begin{array}{l}
-\ell_- (\phi-\phi_-)+V_- \quad\quad \phi<\phi_- \\
\phantom{-}\ell_+(\phi-\phi_-)+V_- \quad \quad\phi>\phi_-
\end{array}
\right.
\end{equation}
and the solutions around the true vacuum are
\begin{equation}
\phi = \mp \frac{\ell_\pm}{2d} \tau^2 + \frac{c_\pm^{(1)}}{\tau^{d-2}} + c_\pm^{(2)}
\end{equation}
where the constants $c_\pm^{(i)}$ are found by requiring $\phi(\tau_0) = \phi_-$ and
$\dot \phi(\tau_0) = \dot \phi_0$.
This implies
\begin{equation}
c_\pm^{(1)} =  \left(\pm \frac{\ell_\mp \tau_0}{d} - \dot \phi_0\right) \frac{\tau_0^{d-1}}{d-2},
\quad
c_\pm^{(2)} = \phi_-+\frac{\dot \phi_0 \tau_0}{d-2} \pm \frac{\ell_\pm \tau_0^2}{2(d-2)}
\end{equation}
The solution $\phi(\tau)$ in the branch $-\ell_-$ passes through the vacuum and climbs 
along the $\ell_+$ branch, until it reaches a maximum. From this maximum  $\phi(\tau)$  comes back along the potential
until the vacuum is  reached again at $\tau = \tau_1$. Requiring $\phi(\tau_1) = \phi_-$
we have
\begin{equation}
\pm
\lambda^\pm
\left(\frac{  \tau _0^2}{d-2}-\frac{  \tau _1^2}{d}\right)+\tau _0 \dot \phi _0=
\frac{2 \tau _0^{d-1} \left(d \phi _0 \pm \ell_\pm  \tau _0\right)}{(d-2) d \tau _1^{d-2}},
\quad 
\dot \phi_0 \tau_1+\dot \phi_1 \tau_0=0
\end{equation}
In this way we can construct the general solution in a stepwise form, by alternating the solution 
along the  $\pm \ell_\pm$ branches of the scalar potential. 
On each branch the solution $\phi(\tau)$ climbs up to a maximum value and then it comes back 
to the vacuum, with non zero derivatives. At this point the solution switches to the other branch of the potential 
reaching another maximum value and coming back.
By requiring continuity of the solution and of its derivative we can express the whole solution in terms of  
$\phi_0$ and of the parameters of the potential.

\begin{figure}
\begin{center}
\begin{tabular}{cc}
\includegraphics[width=7cm]{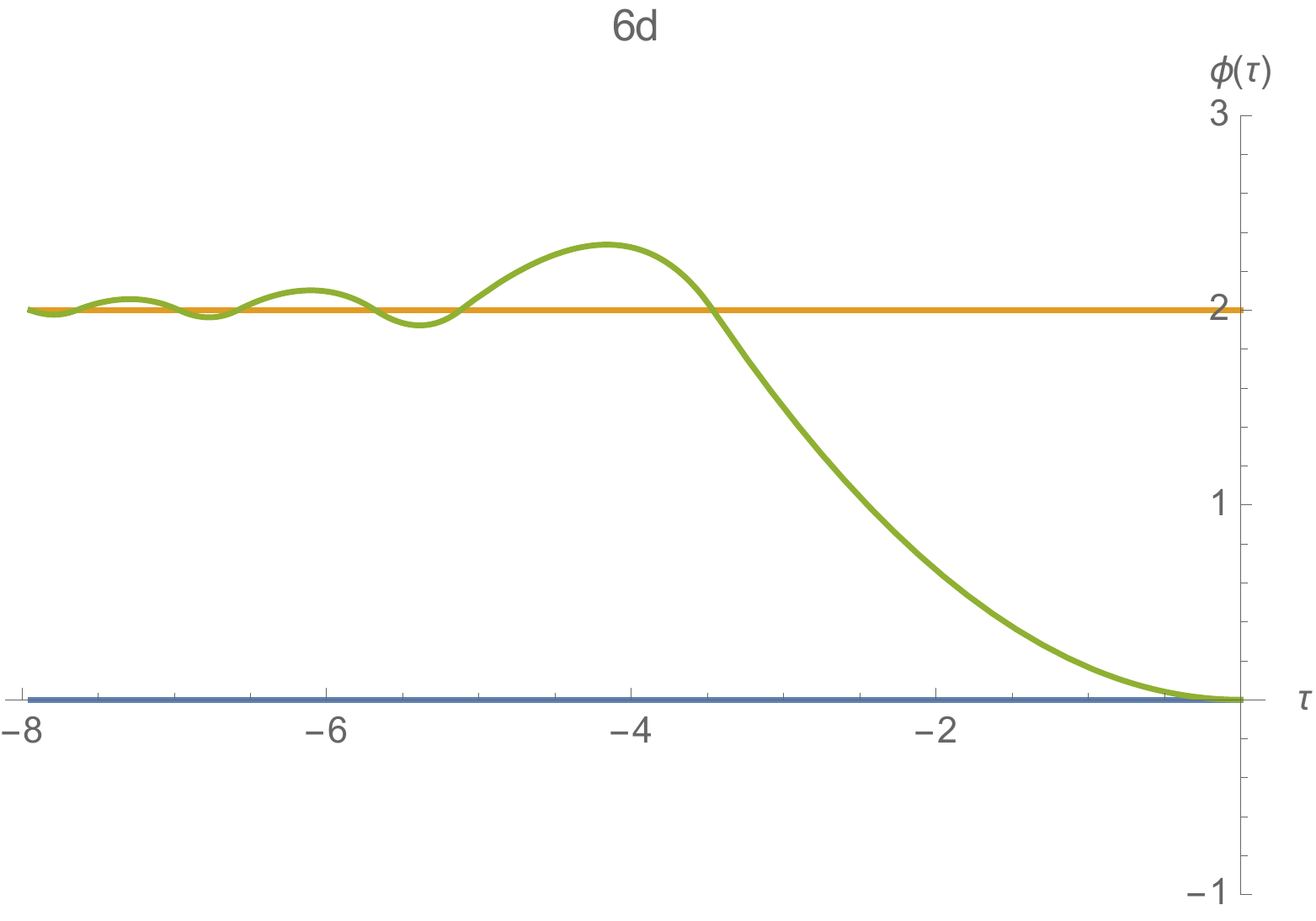}
&
\includegraphics[width=7cm]{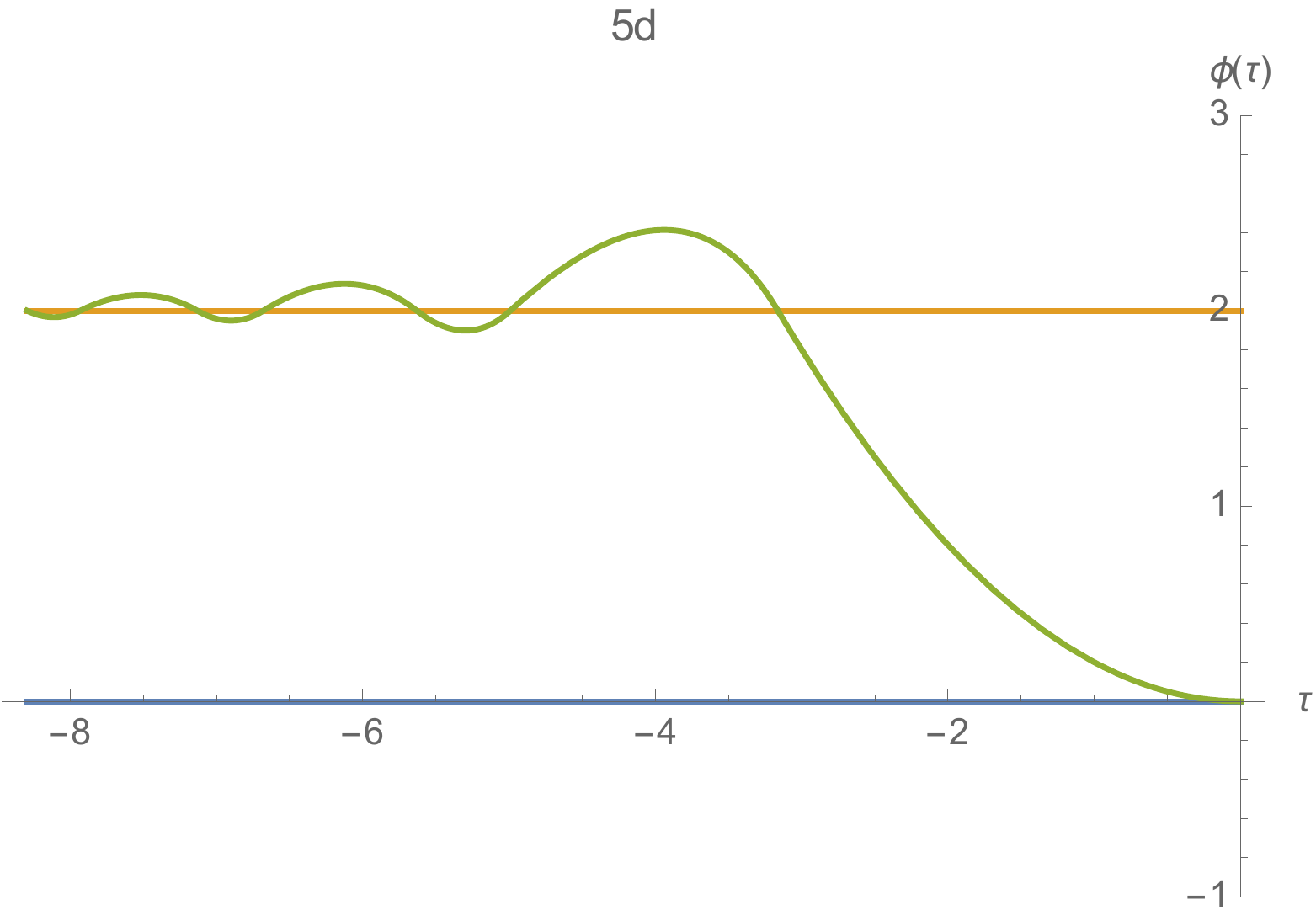}
\\
\includegraphics[width=7cm]{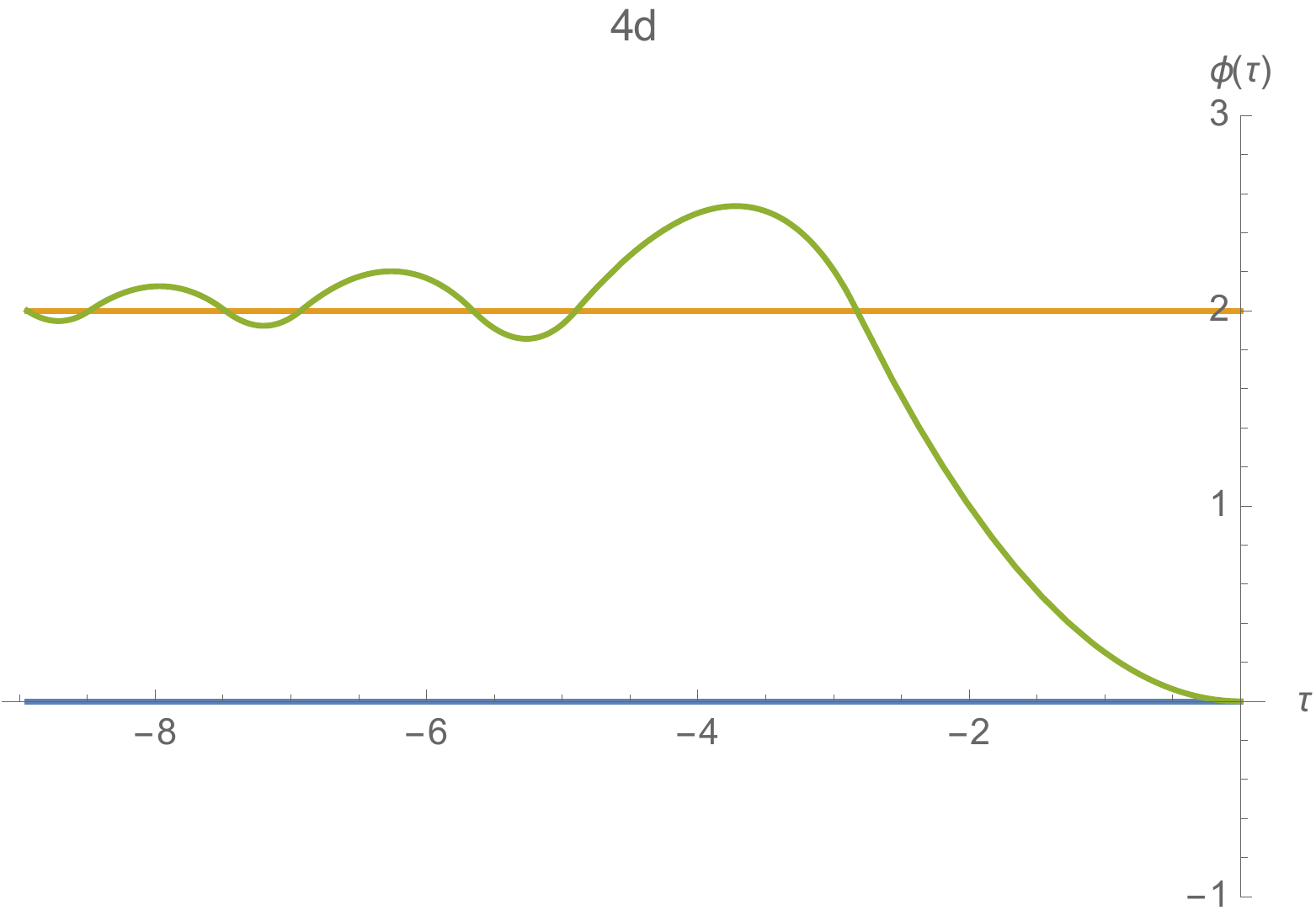}
&
\includegraphics[width=7cm]{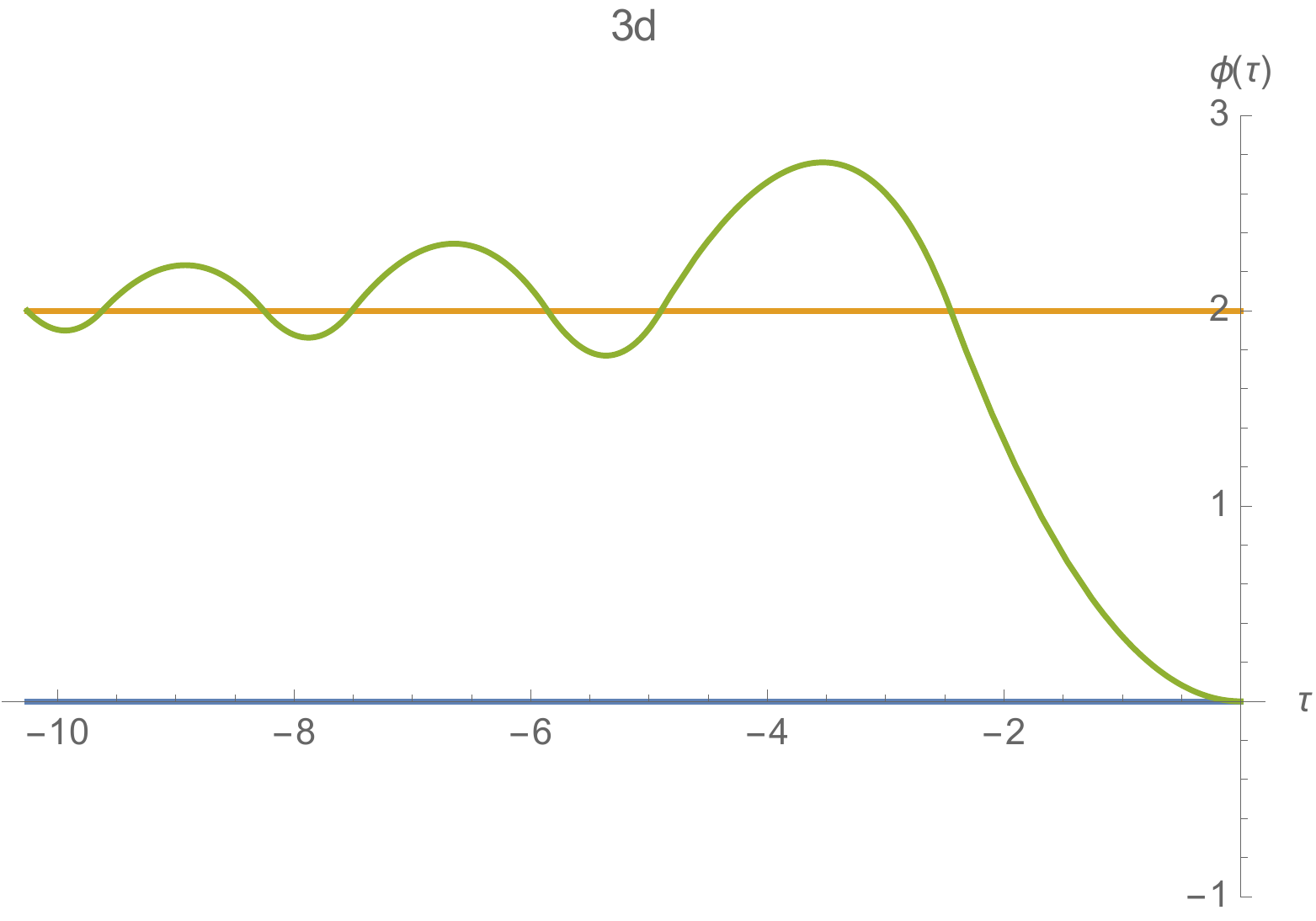}
\end{tabular}
\caption{Damping oscillations around the true vacuum  for the triangular barrier in various dimensions, fixing $\ell_-=2$ and $\ell_+=1$.
The blue line represents the initial value $\phi_0$ and the orange line corresponds to the true vacuum $\phi_-$.}
\label{fig2}
\end{center}
\end{figure}

By following this recursive procedure the solution looks like
\begin{equation}
\phi(\tau)=
\left\{
\begin{array}{ll}
-\frac{\ell_+}{2d} \tau^2+\frac{c_{2n+1}^{(1)}}{\tau^{d-2}} +c_{2n+1}^{(2)}
,\quad\quad & \tau_{2n+1}<\tau<\tau_{2n} \\
\phantom{-}\frac{\ell_-}{2d} \tau^2+\frac{c_{2n}^{(1)}}{\tau^{d-2}} +c_{2n}^{(2)}
,\quad &~~~ \tau_{2n}<\tau<\tau_{2n-1} \\
\end{array}
\right.
\end{equation}
with
\begin{equation}
c_{2n+1}^{(1)} =  -\left( \frac{\ell_{+} \tau_{2n}}{d} + \dot \phi_{2n}\right) \frac{\tau_{2n}^{d-1}}{d-2},
\quad
c_{2n+1}^{(2)} = \phi_-+\frac{\dot \phi_{2n} \tau_{2n}}{d-2} + \frac{\ell_+ \tau_{2n}^2}{2(d-2)}
\end{equation}
and
\begin{equation}
c_{2n}^{(1)} =  \left( \frac{\ell_{-} \tau_{2n-1}}{d} - \dot \phi_{2n-1}\right) \frac{\tau_{2n}^{d-1}}{d-2},
\quad
c_{2n}^{(2)} = \phi_-+\frac{\dot \phi_{2n} \tau_{2n-1}}{d-2} - \frac{\ell_- \tau_{2n-1}^2}{2(d-2)}
\end{equation}
The recurrence relations are
\begin{equation}
\ell_+
\left(\frac{  \tau _{2n-1}^2}{d-2}\!-\!\frac{  \tau _{2n}^2}{d}\right)\! \!+\!\tau _{2n-1} \dot \phi _{2n-1}=
\frac{2 \tau _{2n-1}^{d-1} \left(d \phi _{2n-1} \!+\! \ell_+  \tau_{2n-1}\right)}{d(d-2)  \tau_{2n}^{d-2}},
\,
\dot \phi_{2n-1} \tau_{2n}+\dot \phi_{2n} \tau_{2n-1}\!=\!0
\end{equation}
and
\begin{equation}
-\ell_-
\left(\frac{  \tau _{2n}^2}{d-2}-\frac{  \tau _{2n+1}^2}{d}\right)+\tau _{2n} \dot \phi _{2n}=
\frac{2 \tau _{2n}^{d-1} \left(d \phi _{2n} - \ell_-  \tau_{2n}\right)}{d(d-2)  \tau_{2n+1}^{d-2}},
\,
\dot \phi_{2n} \tau_{2n+1}+\dot \phi_{2n+1} \tau_{2n}=0
\end{equation}
These equation cannot be solved analytically for generic dimension $d$. Nevertheless we can 
study  them numerically by fixing the parameters of the potential and the initial value $\phi_0$.
We have done this analysis for $d=3,\dots,6$ and we have plotted the solution $\phi(\tau)$ 
in Figure \ref{fig2}. From the figure we observe the expected oscillatory damped 
behavior of the solution.

 %
 %
 %
 %
 %
 %%%%%%%%%%%%%%%
 %%%%%%%%%%%%%%%
\section{The square potential}
\label{secsqu}
 %%%%%%%%%%%%%%%
 %%%%%%%%%%%%%%%
 %
 %
 %
 %
 %
In this section we study the tunneling of a false vacuum in $d$-dimensions in presence of a square potential.
This case corresponds to the second picture in Figure \ref{fig0} and it consists of a plateau with $V=V_T$ 
between $\phi_+$ and $\phi_-$. The gradients $\lambda_{\pm}$ discussed in triangular case 
become infinite here.
The solutions $\phi_L(r)$ and $\phi_R(r)$ can be then deduced from (\ref{phiTF})  in this limit.
From (\ref{phiTF}) we have that  $\phi_{\pm} \rightarrow \phi_{T}$ if 
$R_{\pm} \rightarrow R_{T}$ faster than  $\lambda_{\pm}^{-\frac{1}{2}}$.
In this way we have a discontinuity in the derivatives
\begin{equation}
\lim_{R_- \rightarrow R_T } \bigg[ \frac{1}{2} {\phi'_R}^2 - V(\phi) \bigg]_{R_-}^{R_T}=0
,\quad
\lim_{R_+ \rightarrow R_T } \bigg[  \frac{1}{2} {\phi'_L}^2 -V(\phi) \bigg]_{R_T}^{R_+}
\end{equation}
The boundary conditions are then 
\begin{equation}
\label{bcsq1}
\left\{
\begin{array}{lc}
\phi(r) = \phi_-& \quad 0 \leq r \leq R_-  \\
\phi'(r)|_{r=R_-} = -\sqrt{2 \Delta V_-}
\end{array}
\right.
\end{equation} 
on the right side of the barrier and 
\begin{equation}
\label{bcsq2}
\left\{
\begin{array}{l}
\phi(R_+) = \phi_+  \\
\phi'(r)|_{r=R_+} = -\sqrt{2 \Delta V_+}
\end{array}
\right.
\end{equation} 
on the left side of the barrier.
By integrating the equations of motions along the plateau, i.e. in the region $R_-\leq r\leq R_+$, 
we have two equivalent solutions, depending on considering the boundary condition (\ref{bcsq1}) or  (\ref{bcsq2}). These solutions are
\begin{eqnarray}
\label{twosolns}
\phi(r) &=& \phi_-+
\frac{R_- \sqrt {2\Delta V_-} }{d-2} \bigg(\bigg(\frac{R_-}{r} \bigg)^{\!\!d-2} \!\!\!\!\!\!\!-1\bigg)
= \phi_+ +
\frac{R_+ \sqrt {2\Delta V_+} }{d-2} \bigg(\bigg(\frac{R_+}{r} \bigg)^{\!\!d-2} \!\!\!\!\!\!\!-1\bigg)
\end{eqnarray}
The requirement that the two solutions (\ref{twosolns})  coincide gives raise to 
two equations that allow us to express $R_{\pm}$ in terms of the parameters of the potential.
These two equations are obtained matching the terms of order $r^{2-d}$ and $r^0$ in 
(\ref{twosolns}).
We then have the two conditions
\begin{equation}
R_+^{d-1} \sqrt{\Delta V_+} = R_-^{d-1} \sqrt{\Delta V_-}
\end{equation}
and
\begin{equation}
\Delta \phi \equiv \phi_- -\phi_+ = \frac{ \sqrt{2}}{d-2} \left( \sqrt{\Delta V_-}- \sqrt{\Delta V_+} 
\left(\frac{\Delta V_-}{\Delta V_+}\right)^\frac{1}{2(d-1)} \right)
\end{equation}
By integrating the  action and using the various relation we arrive at a simple expression for the
$d$-dimensional bounce
\begin{equation}
\label{Bsq}
B = \frac{\pi^{\frac{d}{2}} (d-2)^{d-1} \Delta \phi^d}{2^{\frac{d-2}{2}} \Gamma \left(1+\frac{d}{2} \right)
\left(\Delta V_-^\frac{d-2}{2(d-1)} -\Delta V_+^\frac{d-2}{2(d-1)} \right)^{d-1}}
\end{equation}
We can also compare this result  with the thin wall approximation.  In this case evaluating $S_1$ in (\ref{s1s1}) 
and plugging it in (\ref{bbbb}) we have
\begin{equation}
\label{twsq}
B_{t.w.} = \frac{(2 \pi)^\frac{d}{2} \Delta V_+^\frac{d}{2} \Delta \phi^d (d-1)^{d-1} }
{\epsilon^{d-1}\Gamma \left(1+\frac{d}{2} \right)}
\end{equation}
It is straightforward to show that (\ref{Bsq}) becomes (\ref{twsq}) if $\Delta V_- = \Delta V_++\epsilon$ in the limit 
$\epsilon \rightarrow 0$.
%
%
%
%
%
%%%%%%%%%%%%%%%%%%%
%%%%%%%%%%%%%%%%%%%
\section{The quadratic approximation}
\label{secqu}
%%%%%%%%%%%%%%%%%%%
%%%%%%%%%%%%%%%%%%%
%
%
%
%
%
In this section we study a more physical situation, by focusing on a scalar potential that is quadratic around both 
the true and the false vacuum.
The four dimensional case has been studied in  detail in \cite{Pastras:2011zr}.
By following the discussion there  we distinguish two possibilities.
In the first case, referred in \cite{Pastras:2011zr} as \emph{volcanic},
the potential has two quadratic branches that are connected 
by a cusp at the local maximum.
In the second case the cusp is replaced by a quadratic cap.

\subsection{The volcanic case}
In this case the scalar potential has two branches, with two quadratic behaviors, both at the true and at the false vacuum.
The mass terms at vacuum with energy $V_\pm$ are denoted as $m_{\pm}$  and the potential is
\begin{equation}
\label{ppo}
V(\phi) = \left \{ 
\begin{array}{cc}
\frac{1}{2} m_+ (\phi-\phi_+)^2 + V_+ & \quad \phi<\phi_T 
\vspace{.1cm}\\
\frac{1}{2} m_- (\phi-\phi_-)^2 + V_- &\quad \phi>\phi_T \\
\end{array}
\right.
\end{equation}
As in the case of the triangular barrier the local maximum is at energy $V_T$, 
and it is obtained by requiring the continuity of the potential $V(\phi)$ at
$\phi = \phi_T$.
In this case the equations of motions assume the form
\begin{equation}
\label{52}
\phi''(r) + \frac{d-1}{r} \phi'(r) = m_{\pm}^2 (\phi(r)-\phi_\pm)
\end{equation}In order to solve them we make the substitution
$\phi(r)-\phi_{\pm} = \frac{y(r)}{r^{\frac{d}{2}-1}}$, such that (\ref{52}) becomes
\begin{equation}
r y''(r) + r^2 y'(r) - \left( m_\pm^2 r^2+\left(\frac{d}{2}-1\right)^2 \right) y(r) = 0 
\end{equation}
These are the modified Bessel equations and they are solved by
\begin{equation}
\phi(r) = \phi_{\pm} + \frac{a_{\pm} I_{\frac{d}{2}-1} (m_\pm r) +b_{\pm} K_{\frac{d}{2}-1} (m_\pm r)  }{r^{\frac{d}{2}-1}}
\end{equation}
The functions $I_{\frac{d}{2}-1}(x)$ and  $K_{\frac{d}{2}-1}(x)$ are the modified Bessel functions
of $\big(\frac{d}{2}-1\big)$-th order of first and second type respectively.
We are now ready to study the solution of the equations of motion that interpolate between the false vacuum and the
true vacuum.

\subsubsection{The interpolating solution}
\label{interpol}
Here we study the solution for the case in which the vacuum cannot be reached at finite Euclidean time. 
In this case the solution is
\begin{equation}
\phi = 
\left\{
\begin{array}{lc}
\phi_{-} + \frac{a_{-} I_{\frac{d}{2}-1} (m_- r) +b_{-}  K_{\frac{d}{2}-1} (m_- r)  }{r^{\frac{d}{2}-1}}
& r<R_T 
\vspace{.1cm} \\
\phi_{+} + \frac{a_{+}  I_{\frac{d}{2}-1} (m_+ r) +b_{+} K_{\frac{d}{2}-1} (m_+ r)  }{r^{\frac{d}{2}-1}}
& R_T<r<R_+ 
\end{array}
\right.
\end{equation}
The condition on the derivative at $r=0$ is
$\phi'(0) = 0$ and it can be imposed by studying the derivative
\begin{equation}
\phi'(r) = \frac{a_{-} I_{\frac{d}{2}} (m_\pm r) -b_{-} K_{\frac{d}{2}} (m_\pm r)  }{r^{\frac{d}{2}-1}}m_{\pm}
\end{equation}
Since  $I_{\frac{d}{2}} (0)=0$ the boundary condition imposes $b_{-} =0$. 
The conditions $\phi(R_+) = \phi_+$ and
$\phi'(R_+) = 0$ impose the two equations
\begin{equation}
\begin{array}{c}
a_{+} I_{\frac{d}{2}-1} (m_+  R_+) + b_{+} K_{\frac{d}{2}-1} (m_+ R_+) =0 
\vspace{.1cm}\\
a_{+}  I_{\frac{d}{2}} (m_+ R_+) - b_{+} K_{\frac{d}{2}} (m_+ R_+) =0 
\end{array}
\end{equation}
The positivity of $I_{n}(x)$ and $K_{n}(x)$ for any $n$ implies that $a_{+}=b_{+} = 0$ 
at finite $R_+$. On the other hand requiring $R_+ \rightarrow \infty$ we have
$K_{\frac{d}{2}} (m_+ R_+) \rightarrow 0$ and we can choose in this case $a_{+} = 0$.
The last condition that we need to solve corresponds to matching the left and the right side of 
the solution at $R_T$.
In this way we can express $a_{-}$ and $b_{+}$ as functions of $R_T$ as
\begin{equation}
a_{-}= \frac{\phi_T-\phi_-}{I_{\frac{d}{2}-1}(m_- R_T)} R_T^{\frac{d}{2}-1}
\quad  \& \quad
b_{+}= \frac{\phi_T-\phi_+}{K_{\frac{d}{2}-1}(m_+ R_T)} R_T^{\frac{d}{2}-1}
\end{equation}
Eventually we obtain an equation for $R_T$ by matching the left and the right derivative of $\phi(r)$ 
at $R_T$
\begin{equation}
\label{RT}
\frac{K_{\frac{d}{2}-1} (m_+ R_T)}{K_{\frac{d}{2}} (m_+ R_T)}
\frac{I_{\frac{d}{2}} (m_- R_T)}{I_{\frac{d}{2}-1} (m_- R_T)}
=\frac{m_+}{m_-} \frac{\phi_T-\phi_+}{\phi_+-\phi_T}
\end{equation}
Summarizing the solution is 
\begin{equation}
\phi = 
\left\{
\begin{array}{lc}
\phi_{-} +(\phi_T-\phi_-) \left(\frac{R_T}{r}\right)^{\frac{d}{2}-1}
\frac{ I_{\frac{d}{2}-1} (m_- r)  }{I_{\frac{d}{2}-1}(m_- R_T)} 
& r<R_T 
\vspace{.1cm} \\
\phi_{+} + (\phi_T-\phi_+) \left(\frac{R_T}{r}\right)^{\frac{d}{2}-1} 
\frac{K_{\frac{d}{2}-1} (m_+ r)}{K_{\frac{d}{2}-1}(m_+ R_T)} 
&r>R_T
\end{array}
\right.
\end{equation}
where $R_T$ is left implicit and it corresponds to the solution of (\ref{RT}).
Computing the bounce action in this case we obtain
\begin{equation}
\label{BBB}
B=
\frac{2 \pi ^{d/2}}{\Gamma \left(\frac{d}{2}\right)}
\left(\frac{m_+ R_T^{d-1} \left(\phi _- -\phi _+ \right) \left(\phi _T-\phi _+\right) K_{\frac{d}{2}}\left(m_+ R_T\right)}{2 K_{\frac{d}{2}-1}\left(m_+ R_T\right)}+\frac{R_T^d \left(V_- -V_+\right)}{d}
\right)
\end{equation}

Using $0<K_\alpha(x)<K_{\alpha+1}(x)$ and $0<J_{\alpha+1}(x) < J_{\alpha}(x)$ 
the relation (\ref{RT}) requires
\begin{equation}
\label{eqfina}
\frac{m_+}{m_-} \frac{\phi_T-\phi_+}{\phi_- -\phi_T} < 1
\end{equation}
If this is not the case we have to look for solutions that reach the true vacuum at finite Euclidean time.
It has been shown in \cite{Pastras:2011zr} that (\ref{eqfina}) is actually always satisfied with the potential given
in formula (\ref{ppo}).
It implies that in the generic $d$-dimensional case the vacuum is never reached at finite Euclidean time.
It is then natural to study the behavior of the solution for Lorentzian time.

\subsubsection{Damped oscillations}

As in the case of the triangular barrier discussed above, when the solution does not reach the true vacuum at finite Euclidean time, we expect that 
it has a damped oscillatory behavior around it, once analytically continued to Lorentzian time.
We then continue the Euclidean time $r$ into the Lorentianz one by the change of variable  $ r \rightarrow i \tau $
and then use the relation $I_\alpha (x) = i^{-\alpha} J(i x)$ between the first 
Bessel function $J(x)$ and the first modified Bessel function $I(x)$.
In this way the solution $\phi(\tau)$ for $r<R_T$ becomes
\begin{equation}
\phi(\tau) = \phi_{-} +(\phi_T-\phi_-) \left(\frac{R_T}{ \tau}\right)^{\frac{d}{2}-1}
\frac{ J_{\frac{d}{2}-1} (m_- \tau)  }{I_{\frac{d}{2}-1}(m_- R_T)} 
\end{equation}
As expected this is an oscillating function and we can study it asymptotically for large $\tau$.
The asymptotic formula is 
\begin{equation}
J_\alpha(x) \rightarrow   \sqrt{\frac{2}{\pi x}} \cos \left(x - \frac{\alpha \pi}{2}-\frac{\pi}{4}\right)
\end{equation}
such that
\begin{equation}
\phi(\tau) \simeq
\phi _--\frac{\left(\phi _--\phi _T\right) 
 \left(\sin \left(\frac{\pi  d}{4}-m_- \tau \right)+\cos \left(\frac{\pi  d}{4}-m_- \tau \right) \right) 
\left(\frac{R_T}{\tau }\right){}^{\frac{d}{2}-1}}{ \sqrt{{\pi  m_- \tau }} \,  I_{\frac{d}{2}-1}\left(m_- R_T\right)}
\end{equation}
where the oscillatory behavior is damped by the $\tau^{-\frac{d}{2}}$
term, generalizing to $d$-dimensions the discussion of \cite{Pastras:2011zr}.

\subsubsection{Recovering the thin wall approximation}

We can compare the bounce action obtained here with the one that one would have obtained from the thin wall approximation in $d$-dimensions.
As a first step it is necessary to compute $\phi_T$ by requiring the continuity of the potential 
at $\phi_T$ for $V_+ = V_- + \epsilon$ or equivalently $\Delta V_- = \Delta V_+ + \epsilon$. 
We have
\begin{equation}
\phi_T = \frac{m_- \phi_- + m_+ \phi_+}{m_-+m_+} + \frac{\epsilon}{m_- m_+ \Delta \phi}
\end{equation}
where $\Delta \phi = \phi_--\phi_+$.
In this way we can estimate the ratio $\frac{m_+ \Delta \phi_+}{m_- \Delta \phi_-} \simeq 1 - \frac{\epsilon \, m_+^2 m_-^2}{(m_++m_-)^2\Delta \phi^2} $,
that implies that $R_T$ has to be large in order to solve equation (\ref{RT}).
It follows that, by using the expansion for the modified Bessel functions at large $R_T$:
\begin{equation}
\label{RT}
\frac{K_{\frac{d}{2}-1} (m_+ R_T)}{K_{\frac{d}{2}} (m_+ R_T)}
\frac{I_{\frac{d}{2}} (m_- R_T)}{I_{\frac{d}{2}-1} (m_- R_T)}
\sim 1-\frac{(d-1) \left(m_++m_-\right)}{2 m_- m_+ R_T}
\end{equation}
In this limit the bounce action (\ref{BBB}) can be approximated as 
\begin{equation}
\label{minetwq}
B= \frac{\left(\pi ^{d/2} \Delta \phi ^{2 d}\right) \left(\frac{d-1}{2 \epsilon }\right)^{d-1} \left(\frac{m_- m_+}{m_++m_-}\right)^d}{d \, \Gamma \left(\frac{d}{2}\right)}
\end{equation}
On the other hand we can compute $S_1$ in (\ref{bbbb}) as in \cite{Pastras:2011zr} and we have
\begin{equation}
\label{s1pas}
S_1 =\frac{ \left(m_- m_+\right)}{2 \left(m_++m_-\right)}\Delta \phi ^2
\end{equation}
Then, by plugging (\ref{s1pas}) into (\ref{bbbb}), we recover (\ref{minetwq}).

\subsection{Smooth quadratic potential}

It is also possible to study a more physical potential, by 
smoothing its shape around the maximum.
For example we can consider the following potential, generalizing the case studied in \cite{Pastras:2011zr}
to $d$-dimensions:
\begin{equation}
V(\phi) = \left \{ 
\begin{array}{cc}
~~\frac{1}{2} m_+^2 (\phi-\phi_+)^2 + V_+, & \quad \phi<\phi_1
\vspace{.1cm}\\
-\frac{1}{2} m_T^2 (\phi-\phi_T)^2 + V_T, & \quad \phi_1<\phi<\phi_2
\vspace{.1cm}\\
~~\frac{1}{2} m_-^2 (\phi-\phi_-)^2 + V_-, &\quad \phi>\phi_2 \\
\end{array}
\right.
\end{equation}
The analysis of the bounce action for such a potential can be performed by distinguishing two cases.
The two cases  differ because the solution can either reach or not $\phi_2$ 
at finite Euclidean time (see Figure \ref{fig0}).

\subsubsection{First case}
In the first case, when $\phi_2$ cannot be reached, the solution of (\ref{eom}) is
\begin{equation}
\label{sol2}
\phi (r) = 
\left\{
\begin{array}{lc}
\phi_{T} + \frac{a_T J_{\frac{d}{2}-1} (m_- r) +b_T Y_{\frac{d}{2}-1} (m_- r)  }{r^{\frac{d}{2}-1}}
& r<R_1
\vspace{.1cm} \\
\phi_{+} + \frac{a_+ I_{\frac{d}{2}-1} (m_+ r) +b_+ K_{\frac{d}{2}-1} (m_+ r)  }{r^{\frac{d}{2}-1}}
& R_1<r<R_+ 
\vspace{.1cm} \\
\phi_+ & r>R_+
\end{array}
\right.
\end{equation}
where in the region $ r< R_1 $ we have substituted the modified Bessel function  $I_\alpha(x)$ and $K_\alpha(x)$
with the Bessel functions $J_\alpha(x)$ and $Y_\alpha(x)$ .
Following the argument in  in section \ref{interpol}, also in this case we have that $R_+\rightarrow \infty$ and the boundary conditions (\ref{bcx}) impose 
\begin{equation}
\label{bc1}
\begin{array}{ccccc}
\lim_{r \rightarrow R_+} \phi(r) &=& \phi_+ &\longrightarrow &\quad a_+ = 0  \\
\frac{\partial \phi(r)}{\partial r} \bigg|_{r=0} &=& 0& \longrightarrow& \quad b_T=0
\end{array}
\end{equation}
The other constants $a_T$ and $b_+$ are determined by requiring that the solution is continuous  at $R_1$
and that $\phi(R_1) = \phi_1$.
These two conditions impose the relations
\begin{equation}
\label{bc2}
a_T= \frac{\phi_1-\phi_T}{J_{\frac{d}{2}-1}(m_T R_1)} R_1^{\frac{d}{2}-1},
\quad  \& \quad
b_+ = \frac{\phi_1-\phi_+}{K_{\frac{d}{2}-1}(m_+ R_1)} R_1^{\frac{d}{2}-1}
\end{equation}
By plugging the constants  obtained in (\ref{bc1}) and (\ref{bc2}) into the relation (\ref{sol2})
the solution becomes 
\begin{equation}
\phi(r) = 
\left\{
\begin{array}{lc}
\phi_{T} -(\phi_T-\phi_1) \left(\frac{R_1}{r}\right)^{\frac{d}{2}-1}
\frac{ J_{\frac{d}{2}-1} (m_T r)  }{J_{\frac{d}{2}-1}(m_T R_1)} 
& r<R_1 
\vspace{.1cm} \\
\phi_{+} + (\phi_1-\phi_+) \left(\frac{R_1}{r}\right)^{\frac{d}{2}-1} 
\frac{K_{\frac{d}{2}-1} (m_+ r)}{K_{\frac{d}{2}-1}(m_+ R_1)} 
&r>R_1
\end{array}
\right.
\end{equation}
The last unknown is $R_1$, and it is obtained by requiring that the solution is smooth at $R_1$
itself. This requirement corresponds to the equation
\begin{equation}
\label{R1}
\frac{K_{\frac{d}{2}-1} (m_+ R_1)}{K_{\frac{d}{2}} (m_+ R_1)}
\frac{J_{\frac{d}{2}} (m_T R_1)}{J_{\frac{d}{2}-1} (m_T R_1)}
=-\frac{m_+}{m_T} \frac{\phi_1-\phi_+}{\phi_T-\phi_1}
\end{equation}
The condition (\ref{R1}) can be used to determine the existence of a solution as well.
As discussed in \cite{Pastras:2011zr} in this case there is an infinite amount of solutions 
allowed by the equations of motion, and we have to select the correct one.
The reason for this infinite amount of solutions is related to the behavior of the function $\frac{J_{\frac{d}{2}}(x)}{J_{\frac{d}{2}-1}(x)}$.
This is a non injective function and it allows for an infinite number of solutions $R_{1}^{(j)}$ with $j=1,\dots,\infty$ for 
(\ref{R1}).

\begin{figure}[htbp]
\begin{center}
\begin{tabular}{cc}
\includegraphics[width=7cm]{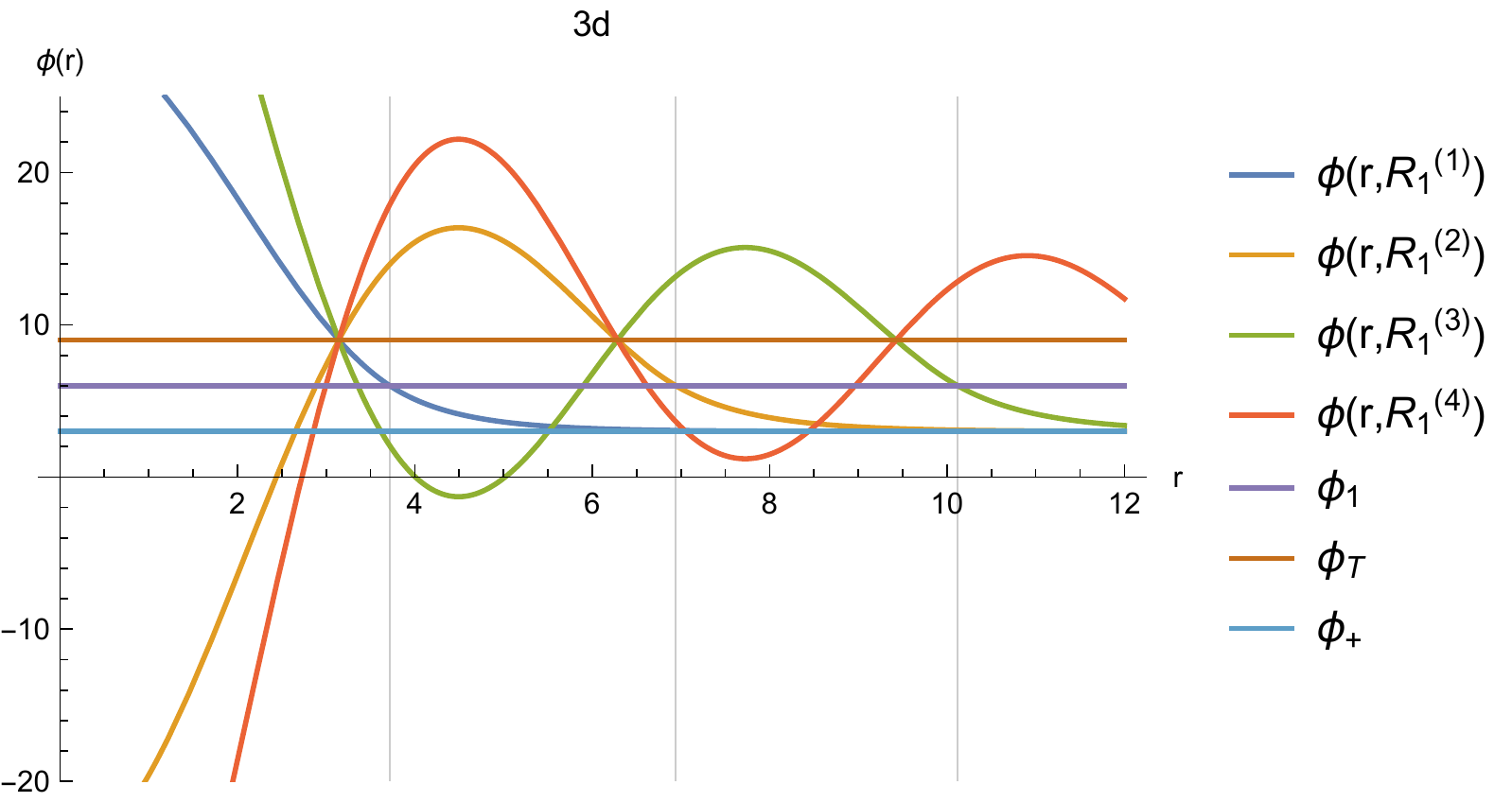}
&
\includegraphics[width=7cm]{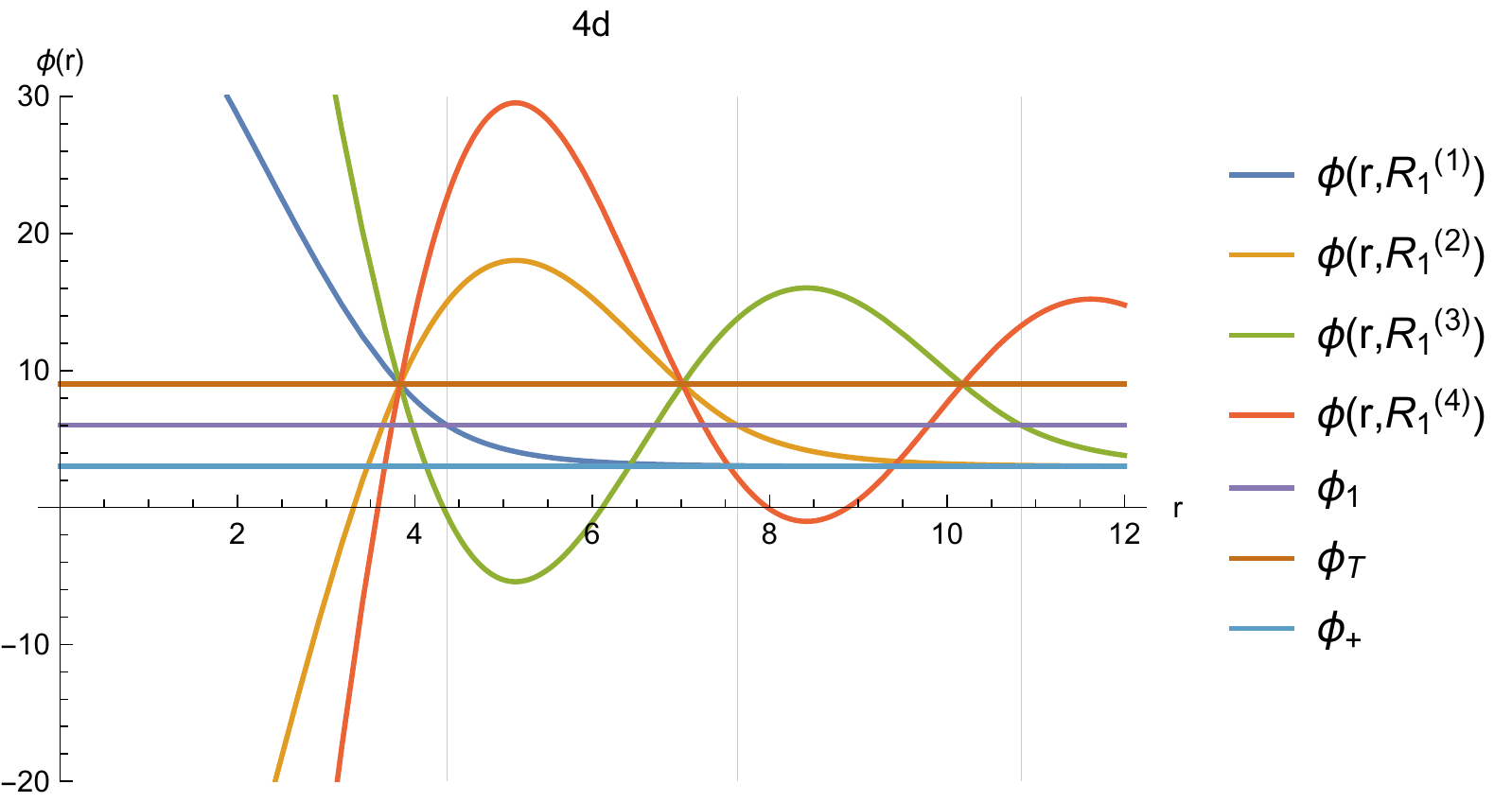}
\\
\includegraphics[width=7cm]{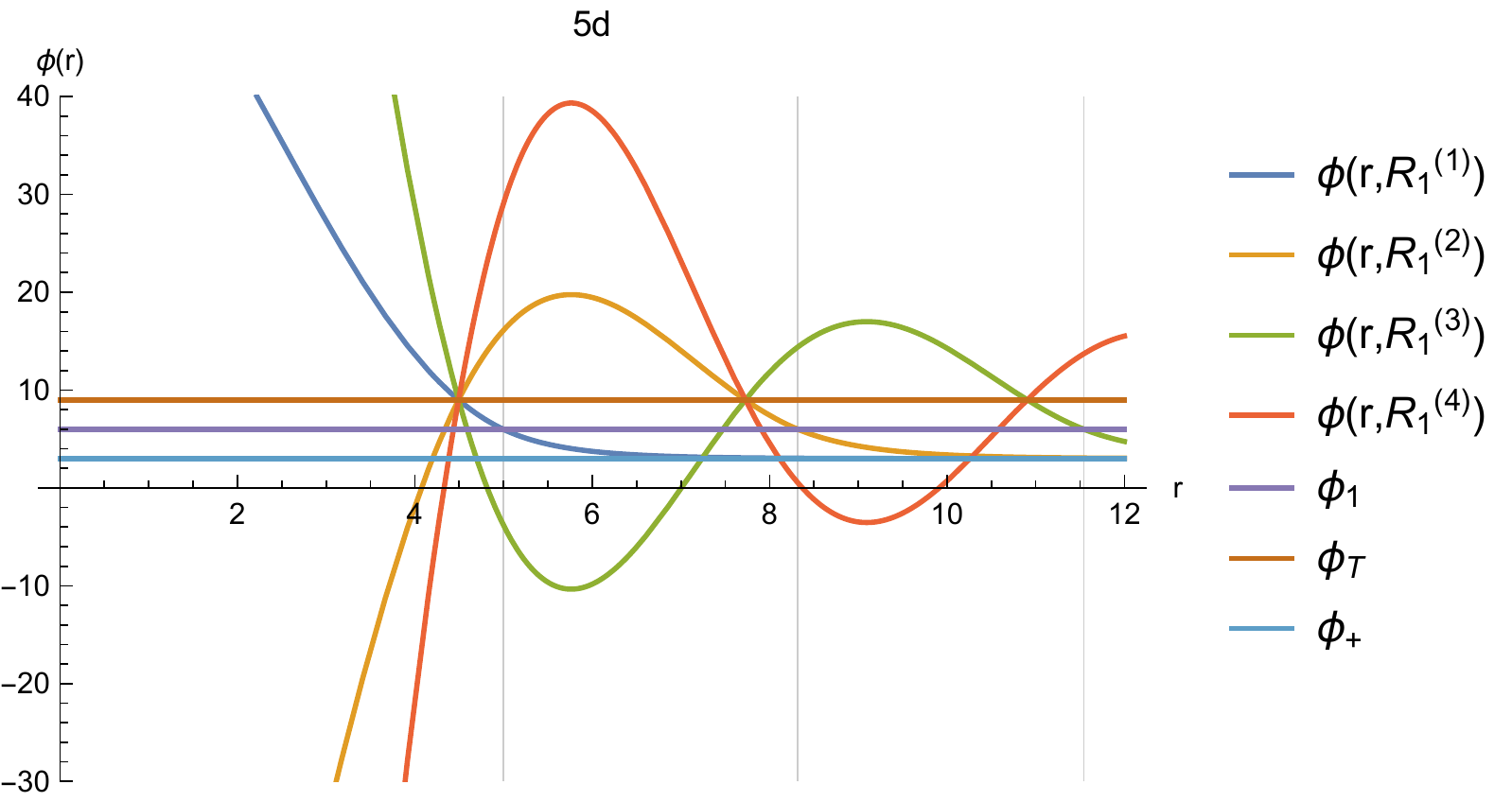}
&
\includegraphics[width=7cm]{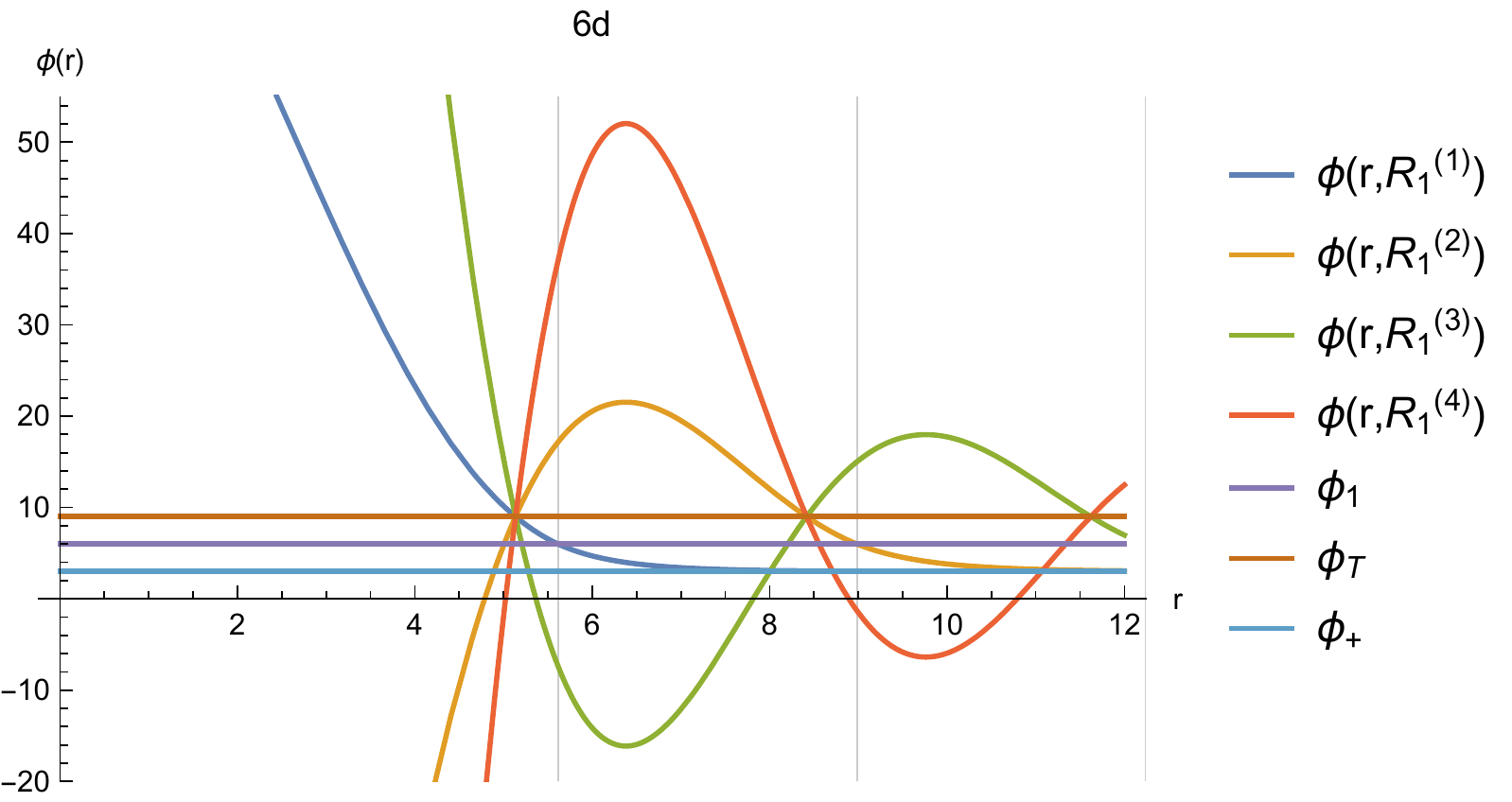}
\end{tabular}
\caption{Various solutions for different values of $R_{1}^{(j)}$ with $j=-1,\dots,4$ in $d=3,\dots,6$. The value of   
 $R_{1}^{(j)}$ is represented by the vertical lines in the figure and it is ordered such that $R_{1}^{(i)} < R_1^{(j)}$ if $i<j$.}
\label{fig4}
\end{center}
\end{figure}

In the following we give a numerical evidence of the fact that there is always one allowed solutions, the one with $R_{1}^{(1)}<R_{1}^{(j\neq 1)}$. 
In Figure \ref{fig4} we plot the various solutions in $d=3,\dots,6$.
From the plots we can see that in the region $r < R_1^{(i)}$, only solution with $r=R_1^{(1)}$ is always such that $\phi(r) < \phi_1$.
Once we have found the correct solution we can use it to compute the bounce action. We obtain
\begin{equation}
S_B
=\frac{ \pi ^{d/2}}{\Gamma \left(\frac{d}{2}\right)}
R_1^d \left(
\frac{2 \left(V_T-V_+\right)}{d}+
(\phi _1-\phi _+)(\phi _T-\phi _+)
\frac{m_+}{R_1} \frac{ K_{\frac{d}{2}}(R_1 m_+)}{K_{\frac{d}{2}-1}(R_1 m_+)}\right)
\end{equation}

On the other hand, we should look for solutions of the second type if $\phi(0)>\phi_2$, i.e. if
\begin{equation}
\label{secondcond}
-\frac{\phi_T-\phi_1}{\phi_2-\phi_T} < \frac{J_{\frac{d}{2}-1} (m_T R_1)}{R_1^{\frac{d}{2}-1}} \Gamma\left(\frac{d}{2}\right) 2^{\frac{d}{2}-1} 
\end{equation}

\subsubsection{Second case}
If the inequality (\ref{secondcond}) is satisfied the solution of (\ref{eom}) with the boundary conditions (\ref{bcx}) is
\begin{equation}
\phi = 
\left\{
\begin{array}{lc}
\phi_- + \frac{a_- I_{\frac{d}{2}-1}(m_- r)}{r^{\frac{d}{2}-1}}
& r<R_2
\vspace{.1cm} \\
\phi_{T} +
\frac{ a_T J_{\frac{d}{2}-1} (m_T r) + b_T Y_{\frac{d}{2}-1} (m_T r)   }{r^{\frac{d}{2}-1}} 
& R_2<r<R_1 
\vspace{.1cm} \\
\phi_{+} + \frac{
a_+
K_{\frac{d}{2}-1} (m_+ r)}{r^{\frac{d}{2}-1}} 
&r>R_1
\end{array}
\right.
\end{equation}
Requiring $\phi(R_1) = \phi_1$ and $\phi(R_2) = \phi_2$, the integration 
constants $a_\pm$ are
\begin{equation}
a_- = \frac{(\phi_2-\phi_-)R_2^{\frac{d}{2}-1}}{I_{\frac{d}{2}-1}(m_- R_2)},
\quad
a_+ = \frac{(\phi_1-\phi_+)R_1^{\frac{d}{2}-1}}{K_{\frac{d}{2}-1}(m_- R_2)}
\end{equation}
From the requirements of continuity at $R_1$ and at $R_2$ we have
\begin{equation}
a_T= -\frac{Y_{\frac{d}{2}-1} (m_T R_2) (\phi_T-\phi_1) R_1^{\frac{d}{2}-1} -Y_{\frac{d}{2}-1} (m_T R_1) (\phi_2-\phi_T) R_2^{\frac{d}{2}-1}}
{J_{\frac{d}{2}-1}(m_T R_1) Y_{\frac{d}{2}-1}(m_T R_2)-Y_{\frac{d}{2}-1}(m_T R_1) J_{\frac{d}{2}-1}(m_T R_2) }
\end{equation}
and
\begin{equation}
b_T = -\frac{J_{\frac{d}{2}-1} (m_T R_2) (\phi_2-\phi_T) R_2^{\frac{d}{2}-1} -J_{\frac{d}{2}-1} (m_T R_2) (\phi_T-\phi_1) R_1^{\frac{d}{2}-1}}
{J_{\frac{d}{2}-1}(m_T R_1) Y_{\frac{d}{2}-1}(m_T R_2)-Y_{\frac{d}{2}-1}(m_T R_1) J_{\frac{d}{2}-1}(m_T R_2) }
\end{equation}
while the smoothness condition at $R_1$ and $R_2$ give two equations for $R_1$ and $R_2$. These are respectively 
\begin{equation}
m_T \left( a_T J_\frac{d}{2} (m_T R_2) + b_T Y_\frac{d}{2}(m_T R_1) \right) = -m_- a_-I_\frac{d}{2}(m_-R_2)
\end{equation}
and
\begin{equation}
m_T \left( a_TJ_\frac{d}{2} (m_T R_1) + b_TY_\frac{d}{2}(m_T R_2) \right) = m_+ a_+ K_\frac{d}{2}(m_+ R_1)
\end{equation}
These equations cannot be solved analytically  and one should study them numerically
to find the interpolating solution. 
%
%
%
%
%%%%%%%%%%%%%%%%
%%%%%%%%%%%%%%%%
\section{Conclusions}
\label{secon}
%%%%%%%%%%%%%%%%
%%%%%%%%%%%%%%%%
%
%
%
%

In this paper we have studied the bounce solution interpolating between the false
and the true vacua of a $d$-dimensional scalar field theory. 
We focused on some specific shapes of the scalar potential,
such to obtain an analytic evaluation of the Euclidean action,
necessary to estimate the decay rate of the false vacuum state. 

In the analysis we have extended most of the results already known in the 4d case to 
the generic $d$-dimensional one. More specifically we have found the instantonic solutions, obtained the bounce actions 
and matched them, when possible, with the result expected from the thin wall approximation. Furthermore we have shown that 
the solutions that do not reach the true vacuum at finite Euclidean  give raise to
damped oscillation around the true vacuum inside the bubble.

Our analysis is semiclassical, and possible quantum corrections are ignored. It should be interesting to 
study them, at least for some of the potentials discussed here in dimensions other than four.
Furthermore it should be possible to introduce gravitational effects in the problem as well, along the lines of 
\cite{Coleman:1980aw}.

Another possible extension to our work consists of finding other classical potentials that allow for an analytic 
analysis in $d$-dimensions.
In general one can study polynomial potentials $V(\phi) \propto \phi^{n+1}$ when the equations of motion can be formulated as
Emden-Fowler equations
\begin{equation}
\label{ELF}
\ddot \phi + \frac{d-1}{r} \dot \phi = \phi^{n}
\end{equation}
For generic values of $n$ and $d$  analytic solutions of these types of equations are not known. Solutions are actually known in the $n=0,1$ case for generic $d$, and these are the cases studied in this 
paper. 
For more general values of $n$ and $d$ only sporadic solutions are known. For example the equation (\ref{ELF}) can be analytically solved in the  $d=3$, $n=6$ - in this case the equation 
 (\ref{ELF}) becomes the Lane–Emden equation -  and in the the $d=6$, $n=3$ cases - see for example \cite{Aslanov2016AnEE} for a solution of this case.
 It is in principle  possible to perform an analytic study  of the bounce solutions in these two cases. 

Let's conclude discussing possible applications of our work. The bounce action for a triangular barrier has already been discussed in three dimensions
\cite{Amariti:2009kb} and the functional behavior was guessed in $d$-dimensions in \cite{Giveon:2009bv}.
The reason behind these result was metastable supersymmetry breaking in generalizations of the ISS model of \cite{Intriligator:2006dd}, when the potential
interpolating between the supersymmetry breaking state and the true SUSY vacuum can be approximated as a triangular barrier.
However one may expect other types of metastable supersymmetric breaking mechanisms other than ISS with other relevant shapes for the scalar potential.

\section*{Acknowledgments}
This work has been supported in part by the Italian Ministero dell'Universit\`a e Ricerca (MUR), 
in part by Istituto Nazionale di Fisica Nucleare (INFN)
through the “Gauge Theories, Strings, Supergravity” (GSS) research project and in
part by MIUR-PRIN contract 2017CC72MK-003.

\appendix

\section{Thin Wall and the triangular barrier in $d=3,4$ and $6$}
\label{app36}

In this appendix we study the limit of the bounce action for $\Delta V_- - \Delta V_+ = \epsilon \rightarrow 0$ in the case of the 
triangular potential in $4$, $3$ and $6$ dimensions.
From these result we have guessed in section \ref{secondTri} the general behavior in $d$ dimensions and compared it with the thin wall approximation

\subsection{3d}
\label{app3}
In three dimensions we can solve the  equations for $\Delta \phi_\pm$  in (\ref{phiTF}). The relevant solution suitable for a large $R_T$ expansion
in the thin wall limit is
\begin{eqnarray}
\label{sol3d}
R_+=\frac{1}{2} \left(R_T+\frac{\sqrt[3]{\Xi_+}}{\lambda _+}+\frac{\lambda _+ R_T^2}{\sqrt[3]{\Xi_+}}\right)
,\quad
R_-=\frac{1}{2} \left(R_T-\frac{ e^{\frac{i \pi}{3} }  \sqrt[3]{\Xi_-}}{\lambda _-}-\frac{ \lambda _- R_T^2}{e^{\frac{i \pi}{3}}  \sqrt[3]{\Xi_-}}\right)
\end{eqnarray}
\begin{equation}
\Xi_\pm = \Theta _\pm +12 \Delta \phi _\pm  \lambda _\pm ^2 R_T-\lambda _\pm^3 R_T^3,
\quad
\Theta_{\pm} = 
2  \lambda _\pm^2 R_T \sqrt{6  \Delta \phi _\pm \left(6 \Delta \phi _\pm^2 - \lambda _\pm R_T^2\right)}
\end{equation}
We then study the limit $\Delta V_- - \Delta V_+ = \epsilon \rightarrow 0$. This is done solving the first equation
in  (\ref{phiTF})  by expanding  $R_T$ as $\frac{\alpha}{\epsilon} + \beta$.
We obtain $\alpha = \frac{4 \Delta \phi \sqrt{2 \Delta V_+} }{3} $
while we do not solve for $\beta$. Indeed its actual value is irrelevant for the limit of the bounce action
that  we are going to compute.
Plugging $R_T$ expanded at such order in (\ref{sol3d}) the corresponding expressions for $R_\pm$ are
\begin{eqnarray}
R_\pm = \frac{4 \Delta \phi \sqrt{2\Delta V_+}  }{ 3 \epsilon }\pm\frac{\sqrt{2} \Delta \phi_\pm} {\sqrt{\Delta V_+}}+\beta+O(\epsilon )
\end{eqnarray}
In this way we obtain for the bounce action
\begin{equation}
B \simeq
\frac{256 \sqrt{2} \pi  \Delta V_+^{3/2} \Delta \phi ^3}{81 \epsilon ^2}
\end{equation}
that coincides with the result expected from the thin wall approximation. Observe that at this order there is no $\beta$ dependence 
in the bounce action: this is the reason why we did not compute $\beta$ explicitly.

\subsection{4d}

In 4d the limit of the  exact bounce action in the regime $\Delta V_- = \Delta V_+ + \epsilon  + o(\epsilon^2) $ and 
the relation with the result expected from the thin wall approximation has been already performed in \cite{Duncan:1992ai}.
However here we analyze this case using the same logic discussed in appendix \ref{app3} such to provide an unified 
formalism and to give evidence of the guess done in equation (\ref{thesew}).
In this case the equation  (\ref{phiTF})  is solved by
\begin{equation}
\label{sol4d}
R_\pm=\sqrt{R_T \left(R_T \pm \sqrt{\frac{8 \Delta \phi _\pm^2}{\Delta V_\pm}}  \right)}
\end{equation}
We then study the limit $\Delta V_- - \Delta V_+ = \epsilon \rightarrow 0$. This is done solving the system of equations
(\ref{eq6d})  by expanding  $R_T$ as $\frac{\alpha}{\epsilon} + \beta$.
We obtain $\alpha = 2\Delta \phi \sqrt{2 \Delta V_+} $
while we do not solve for $\beta$. 
Even if in this case the equations can be solved analytically  here we keep the same perturbative approach 
used in the other case, such to have a uniform description.
Again we will see that the value of $\beta$  is irrelevant when 
we compare our result with the one obtained from the 
thin wall approximation.
Plugging $R_T$ expanded at such order in (\ref{sol4d}) the corresponding expressions for $R_\pm$ are
\begin{equation}
R_\pm = \frac{2 \Delta \phi \sqrt{2\Delta V_+}  }{ \epsilon }\pm\frac{\sqrt{2} \Delta \phi_\pm} {\sqrt{\Delta V_+}}+\beta+O(\epsilon )
\end{equation}
Plugging these values in the bounce action and expanding at small $\epsilon$ the leading term is
\begin{equation}
B \simeq
\frac{32 \pi ^2   \Delta  V_-^2 \Delta \phi ^4}{3 \epsilon ^3}
\end{equation}
that does not depend on $\beta$ and coincides with the result expected from the thin wall approximation. 

\subsection{6d}

In the six dimensional  case the last two equations in (\ref{phiTF}) become
\begin{equation}
\label{eq6d}
\Delta  \phi _\pm=\frac{\lambda _\pm \left(R_\pm^2-R_T^2\right)^2 \left(2 R_T^2+R_\pm^2\right)}{24 R_T^4}
\end{equation}
i.e. we have a  cubic system of equation  in $R_{\pm}^2$ and we can analytically solve it. The relevant solution is
\begin{equation}
\label{sol6d}
R_+=\sqrt{\frac{\lambda _+^2 R_T^4+\Theta _+^{2/3}}{\sqrt[3]{\Theta _+} \lambda _+}},
\quad
R_-=\sqrt{\frac{\lambda _-^2 R_T^4+e^{\frac{2 i \pi }{3}} \Theta _-^{2/3}}{e^{\frac{i \pi }{3}} \sqrt[3]{\Theta _-} \lambda _-}}
\end{equation}
with
\begin{equation}
\Theta _\pm=\lambda _\pm^2 R_T^4 \left(12 \Delta \phi _\pm+2 \sqrt{6 \Delta \phi _\pm \left(6 \Delta \phi _\pm-\lambda _+ R_T^2\right)}-\lambda _\pm R_T^2\right)
\end{equation}
We then study the limit $\Delta V_- - \Delta V_+ = \epsilon \rightarrow 0$. This is done solving the system of equations
(\ref{eq6d})  by expanding  $R_T$ as $\frac{\alpha}{\epsilon} + \beta$.
We obtain $\alpha = \frac{10 \Delta \phi \sqrt{2 \Delta V_+} }{3} $
while we do not solve for $\beta$. As before we will see that indeed the value of $\beta$  is irrelevant when 
we compare our result with the one obtained from the 
thin wall approximation.
Plugging the expanded value of $R_T$  in (\ref{sol6d}) the corresponding expressions for $R_\pm$ are
\begin{equation}
R_\pm = \frac{10 \Delta \phi \sqrt{2\Delta V_+}  }{3 \epsilon }\pm\frac{\sqrt{2} \Delta \phi_\pm} {\sqrt{\Delta V_+}}+\beta+O(\epsilon )
\end{equation}
In this way we obtain for the bounce action
\begin{equation}
B \simeq
\frac{2^9 \, 5^6 \pi ^3   \Delta  V_+^3 \Delta\phi ^6}{3^7 \, \epsilon ^5}
\end{equation}
that coincides with the result expected from the thin wall approximation. Observe again the absence of $\beta$ in the final result.

\bibliographystyle{JHEP}
\bibliography{ref}

\end{document}